\documentclass[nofootinbib,reprint,amsmath,amssymb,prd,aps,superscriptaddress]{revtex4-1}
\usepackage{graphicx}
\usepackage[T1]{fontenc}
\usepackage{dcolumn}
\usepackage{xcolor,colortbl}
\usepackage{multirow}
\usepackage{comment}
\usepackage{enumitem}
\usepackage[caption=false]{subfig}
\usepackage{bm}

\usepackage{hyperref}
\hypersetup{colorlinks, linkcolor={teal},citecolor={blue},urlcolor={blue}}

\newcommand{\dd}{{\rm d}}

\begin{document}

\title{Black hole solutions in the revised Deser-Woodard nonlocal theory of gravity}

\author{Rocco D'Agostino}
\email{rocco.dagostino@inaf.it}
\affiliation{INAF -- Osservatorio Astronomico di Roma, Via Frascati 33, 00078 Monte Porzio Catone, Italy}
\affiliation{INFN -- Sezione di Roma 1, P.le Aldo Moro 2, 00185 Roma, Italy}

\author{Vittorio~De~Falco}
\email{v.defalco@ssmeridionale.it}
\affiliation{Scuola Superiore Meridionale,  Largo San Marcellino 10, 80138 Napoli, Italy}
\affiliation{INFN -- Sezione di Napoli, Via Cintia, 80126 Napoli, Italy}

\begin{abstract}
We consider the revised Deser-Woodard model of nonlocal gravity by reformulating the related field equations within a suitable tetrad frame. This transformation significantly simplifies the treatment of the ensuing differential problem while preserving the characteristics of the original gravitational theory. We then focus on static and spherically symmetric spacetimes in vacuum. 
Hence, we demonstrate that the gravitational theory under study admits a class of black hole solutions characterized by an inverse power-law correction to the Schwarzschild $g_{tt}$ metric function and a first-order perturbation of the $g_{rr}$ Schwarzschild component. Then, through a stepwise methodology, we analytically solve the full dynamics of the theory, finally leading to the reconstruction of the nonlocal distortion function, within which the new black hole solutions arise. Furthermore, we analyze the geometric properties of the obtained solutions and quantify the deviations from the Schwarzschild prediction. This work provides new insights into compact object configurations and advances our understanding of nonlocal gravity theories in the strong-field regime at astrophysical scales. 
\end{abstract}

\maketitle

\section{Introduction}
Developing a unified theory of gravity that reconciles classical and quantum frameworks remains one of the most significant theoretical challenges in modern physics. While general relativity (GR) elegantly explains gravity as the curvature of spacetime resulting from mass and energy, it lacks a quantum formulation that aligns with the other fundamental forces \cite{Padmanabhan:2001ev,Ashtekar:2004eh}. On the one hand, GR has been extensively validated through various solid tests \cite{Will:2014kxa,LIGOScientific:2016aoc,EventHorizonTelescope:2019dse}; however, some theoretical shortcomings present at different scales  \cite{Ishak:2018his,Barack:2018yly,DAgostino:2023cgx} impose limitations that become particularly significant in extreme gravity regimes, such as in the vicinity of black holes (BHs).

Also, the standard cosmological model, grounded in Einstein's gravity, depicts a Universe predominantly composed of enigmatic and poorly understood components, such as dark energy and dark matter  \cite{SupernovaSearchTeam:1998fmf,SupernovaCosmologyProject:1998vns,Peebles:2002gy,Planck:2018vyg,DAgostino:2019wko}. Notably, the cosmological constant, which drives the late-time Universe's expansion, presents a major conflict when interpreted as vacuum energy predicted by quantum field theory \cite{Weinberg:1988cp,Carroll:2000fy,DAgostino:2022fcx}.
This fundamental tension indicates that the current cosmological paradigm may not be the conclusive picture to describe the observed large-scale structures of the Universe. Therefore, numerous alternative scenarios have been investigated over the past two decades, with the aim of providing a more comprehensive understanding of the interplay between gravity and quantum mechanics \cite{Silvestri:2009hh,DeFelice:2010aj,Clifton:2011jh,Joyce:2014kja,Nojiri:2017ncd,Capozziello:2019cav,DAgostino:2019hvh,DAgostino:2022tdk,DAgostino:2024ymo}.

One of the promising approaches to handling the challenges outlined above involves modifying the gravitational sector by introducing nonlocal terms, which account for the influence of the entire spacetime. In fact, nonlocal models have demonstrated their capability to address BH singularities, but also to explain inflationary dynamics, dark energy, and formation of cosmic structures \cite{Arkani-Hamed:2002ukf,Nojiri:2007uq,Calcagni:2007ru,Biswas:2011ar,Park:2012cp,Maggiore:2014sia,Dirian:2014ara,Capozziello:2022rac,Deffayet:2024ciu}. By relaxing the classical locality, these theories mitigate the instabilities often associated with higher-order derivative operators in ultraviolet extensions of GR. This leads to renormalizable Lagrangians and naturally incorporates the nonlocal terms that arise from loop corrections, making them a potentially significant step toward the formulation of a complete and final theory of quantum gravity \cite{Modesto:2011kw,Modesto:2017sdr,Buoninfante:2018mre}.

A well-known example of a nonlocal gravity scenario is the Deser-Woodard model \cite{Deser:2007jk}. This framework was proposed to reproduce the late-time history of the Universe without the fine-tuning issues of the cosmological constant. Nevertheless, the original model was found to fail in predicting small-scale observations due to the lack of a suitable screening mechanism in the solar system \cite{Belgacem:2018wtb}. The same authors subsequently proposed a refined model to address this limitation \cite{Deser:2019lmm}. This improved version has been applied to a variety of contexts, including cosmology \cite{Ding:2019rlp,Chen:2019wlu,Jackson:2021mgw,Capozziello:2023ccw,Jackson:2023faq} and BH quasi-normal modes \cite{Chen:2021pxd}.

Despite the considerable theoretical progress, the high complexity, typical of most extended gravity theories, severely limits the search for new astrophysical solutions in nonlocal gravity studies. For the same reason, perturbative and/or numerical techniques are often needed to describe the physics of compact objects. Specifically, static and spherically symmetric solutions arising from a massive nonlocal modification of GR at the infrared scales were investigated in \cite{Kehagias:2014sda}. Also, the stability of linear perturbations for weakly nonlocal theories with a finite behavior at the quantum level was examined in \cite{Calcagni:2018pro}. Furthermore, the Newmann-Janis algorithm was used to perturbatively solve the nonlocal field equations for a slowly rotating BH \cite{Kumar:2019uwi}. More recently, a status report on BH-like solutions arising from nonlocal gravity theories endowed with at least quadratic curvature Lagrangians has been presented in  \cite{Buoninfante:2022ild}.

To alleviate the challenging task in the resolution of the nonlocal field equations, we propose a new approach to reformulate the Deser-Woodard differential problem in a suitable tetrad frame. In doing so, we identify new solutions corresponding to static and spherically symmetric BHs and compare their physical predictions with those of the Schwarzschild spacetime. 

The paper is organized as follows. In Sec.~\ref{sec:nonlocal-theory}, we recap the main aspects of the Deser-Woodard model of nonlocal gravity and recast the field equations in a proper tetrad frame simplifying the handling of the differential problem. In Sec.~\ref{sec:BH-solutions}, we describe the strategy to solve the theory and determine static and spherically symmetric BH solutions. In Sec.~\ref{sec:BH-properties}, we discuss the main geometrical properties of the obtained solutions. Finally, we provide conclusions and future perspectives in Sec.~\ref{sec:end}.

Throughout this paper, we adopt units of $c=\hbar=1$. Moreover, we assume $G=M=1$, where $G$ and $M$ are the gravitational constant and the BH mass, respectively. The flat metric is indicated by $\eta_{\mu \nu }={\rm diag}(-1,1,1,1)$.

\section{Deser-Woodard nonlocal gravity in the proper tetrad frame}
\label{sec:nonlocal-theory}
We start by recalling the action of the Deser-Woodard nonlocal gravity model \cite{Deser:2019lmm}:
\begin{equation} \label{eq:nonlocal-action}
S=\dfrac{1}{16\pi}\int \dd^4 x \sqrt{-g}\, R\left[1+f(Y)\right]\,,
\end{equation}
where $g$ is the determinant of the metric tensor $g_{\mu \nu}$, $R$ is the Ricci scalar, and the distortion function $f(Y)$ accounts for the nonlocal gravity effects. The localized form of the action \eqref{eq:nonlocal-action} can be defined in terms of four independent auxiliary scalar fields $\{X,Y,U,V\}$, whose dynamics are governed by the Klein-Gordon equations
\begin{subequations}
\begin{align}
    \Box X&=R\,, \label{eq:X} \\ 
    \Box Y&=g^{\mu\nu}\partial_\mu X\partial_\nu X\,, \label{eq:Y}\\
    \Box U&=-2\nabla_\mu (V\nabla^\mu X)\,, \label{eq:U} \\
    \Box V&=R\frac{\dd f}{\dd Y}\,.\label{eq:V}
\end{align}
\end{subequations}
Here, $\Box\equiv \nabla_\mu \nabla^\mu$ is the relativistic d'Alembert operator, which can be defined on a generic function $u$ as
\begin{equation} 
    \Box u\equiv\frac{1}{\sqrt{-g}}\partial_\alpha\left[\sqrt{-g}\,\partial^\alpha u\right].
    \label{eq:box-operator}
\end{equation}

The vacuum field equations can then be obtained by varying the action \eqref{eq:box-operator} with respect to the metric \cite{Deser:2019lmm}:
\begin{align}
&\left(G_{\mu\nu}+g_{\mu\nu}\Box-\nabla_\mu \nabla_\nu\right)\left[1+U+f(Y)\right]+\mathcal{K}_{(\mu\nu)}\notag  \\
&-\frac{1}{2}g_{\mu\nu}g^{\alpha\beta}\mathcal{K}_{\alpha\beta} =0\,,
\label{eq:FE}
\end{align}
where $G_{\mu\nu}\equiv R_{\mu\nu}-\frac{1}{2}g_{\mu\nu}R$ is the Einstein tensor and $R_{\mu\nu}$ is the Riemann tensor. Here, $\mathcal{K}_{(\mu\nu)}\equiv \frac{1}{2}(\mathcal{K}_{\mu\nu}+\mathcal{K}_{\nu\mu})$, with
\begin{equation}\label{eq:B}
\mathcal{K}_{\mu\nu}:= \partial_\mu X\partial_\nu U +\partial_\mu Y \partial_\nu V+V\partial_\mu X \partial_\nu X \,.
\end{equation}

Let us consider a generic static and spherically symmetric metric:
\begin{equation}
\dd s^2=g_{tt}(r) \dd t^2+g_{rr}(r)\dd r^2 +r^2(\dd\theta^2+\sin^2\theta\, \dd\varphi^2)\,.
    \label{eq:general-metric}
\end{equation}
We proceed by taking into account the tetrad field associated with a static observer at infinity, given by $\left\{{\bf e}_t\,, {\bf e}_r\,,{\bf e}_\theta\,,{\bf e}_\varphi\right\}=\left\{\partial_t\,,\partial_r\,,\partial_\theta,\partial_\varphi\right\}$ and that related to a static observer in the spacetime \eqref{eq:general-metric}, namely \cite{Morris:1988cz}
\begin{align}
{\bf e}_{\hat t} = \frac{{\bf e}_t}{\sqrt{-g_{tt}}},\ 
{\bf e}_{\hat r}= \frac{{\bf e}_r}{\sqrt{g_{rr}}},\
{\bf e}_{\hat \theta} =\frac{{\bf e}_\theta}{r},\ 
{\bf e}_{\hat \varphi}=\frac{{\bf e}_\varphi}{r\sin\theta},
\label{eq:proper_tetrad}
\end{align}
such that $g_{\hat\mu\hat\nu}=e_{\hat \mu}^\alpha e_{\hat \nu}^\beta g_{\alpha\beta}\equiv\eta_{\mu\nu}$, where
\begin{equation}\label{eq:general-tetrad}
    e^{\hat \alpha}_{\mu}:=\text{diag}\left(\frac{1}{\sqrt{-g_{tt}}},\frac{1}{\sqrt{g_{rr}}},\frac{1}{r},\frac{1}{r\sin\theta}\right),
\end{equation}
with the inverse tetrad being
\begin{equation}\label{eq:general-tetrad-inverse}
    e_{\hat \alpha}^{\mu}:=\text{diag}\left(\sqrt{-g_{tt}},\sqrt{g_{rr}},r,r\sin\theta\right).
\end{equation}
In view of this, the Riemann tensor, Ricci tensor, scalar curvature, and Einstein tensor transform, respectively,
\begin{subequations}
\begin{align}
R^{\hat \alpha}{}_{{\hat \beta}{\hat \gamma}{\hat \delta}} &= e^{\hat \alpha}_{\mu} e_{\hat \beta}^{\nu} e_{\hat \gamma}^{\rho} e_{\hat \delta}^{\sigma} R^\mu{}_{\nu \rho \sigma}\,,\\
R_{\hat\mu\hat\nu}&=R^{\hat\alpha}_{\ \hat\mu\hat\alpha\hat\nu}\,,\\  R&=\eta^{\hat\mu\hat\nu}R_{\hat\mu\hat\nu}\,,\\
G_{\hat\mu\hat\nu}&=R_{\hat\mu\hat\nu}-\frac{1}{2}\eta_{\mu\nu} R\,.
\end{align}    
\end{subequations}
Thus, after introducing the function
\begin{equation}\label{eq:W}
W(r):= 1+U(r)+f(Y(r))\,,   
\end{equation}
we recast Eq.~\eqref{eq:FE} as
\begin{align}
&\left(G_{\hat\mu\hat\nu}+\eta_{\mu\nu}\Box-\nabla_{\hat\mu} \nabla_{\hat\nu}\right)W
+\mathcal{K}_{(\hat\mu\hat\nu)}-\frac{1}{2}\eta_{\mu\nu}\eta^{\alpha\beta}\mathcal{K}_{\hat\alpha\hat\beta} =0\,,
\label{eq:Local_FE}
\end{align}
where $\nabla_{\hat\mu} \nabla_{\hat\nu}=e^\alpha_{\hat\mu}e^\beta_{\hat\nu}\nabla_\alpha \nabla_\beta$, $\mathcal{K}_{\hat\mu\hat\nu}=e^\alpha_{\hat\mu}e^\beta_{\hat\nu}\mathcal{K}_{\alpha\beta}$, and the scalar $\Box W$ being calculated as in Eq.~\eqref{eq:box-operator}, since it is an invariant quantity on the same level of $R$.

The non-vanishing diagonal components of Eq.~\eqref{eq:Local_FE} are
\begin{subequations}\label{eq:FE_LOC}
\begin{align}
(G_{\hat t\hat t}-\Box -\nabla_{\hat t}\nabla_{\hat t})W+\frac{1}{2}\mathcal{K}_{\hat r\hat r}&=0,\label{eq:FE_LOC_t}\\    
(G_{\hat r\hat r}+\Box -\nabla_{\hat r}\nabla_{\hat r})W+\frac{1}{2}\mathcal{K}_{\hat r\hat r}&=0,\label{eq:FE_LOC_r}\\ 
(G_{\hat \varphi\hat \varphi}+\Box -\nabla_{\hat \varphi}\nabla_{\hat \varphi})W-\frac{1}{2}\mathcal{K}_{\hat r\hat r}&=0.\label{eq:FE_LOC_phi}    
\end{align}
\end{subequations}
Combining Eqs.~\eqref{eq:FE_LOC_t}, \eqref{eq:FE_LOC_r}, and  \eqref{eq:FE_LOC_phi}, we find the following independent equations:
\begin{subequations}\label{eq:LOC_SUM}
\begin{align}
    &(G_{\hat t\hat t}+G_{\hat\varphi\hat\varphi})W=(\nabla_{\hat t}\nabla_{\hat t}+\nabla_{\hat\varphi}\nabla_{\hat\varphi})W\,,\label{eq:LOC_SUM1} \\
    &(G_{\hat r\hat r}+G_{\hat\varphi\hat\varphi})W+2\Box W=(\nabla_{\hat r}\nabla_{\hat r}+\nabla_{\hat\varphi}\nabla_{\hat\varphi})W\,.\label{eq:LOC_SUM2}
\end{align}
\end{subequations}
We note that the above differential problem is easier to handle compared to Eq.~\eqref{eq:FE}, see Appendix \ref{sec:full}. In the next paragraph, we shall show a method to solve them.

\section{Static and spherically symmetric black holes}
\label{sec:BH-solutions}

To solve the differential problem \eqref{eq:LOC_SUM} and find the distortion function of the theory, we need first to specify the metric components $g_{tt}$ and $g_{rr}$. Specifically, we look for BH solutions arising from Eq.~\eqref{eq:general-metric} with
\begin{equation} \label{eq:BH-metric}
    g_{tt}=-A(r)\,, \qquad g_{rr}=B(r)^{-1}\,,
\end{equation}
where $A(r)$ and $B(r)$ are unknown functions that must satisfy the following conditions:
\begin{enumerate}[label=(\roman*)]
    \item the metric is asymptotically flat, namely 
\begin{equation}\label{eq:AF}
\lim_{r \to \infty}A(r)=\lim_{r \to \infty}B(r)=1\,;
\end{equation} 
    \item they both feature an event horizon (or apparent singularity), $r_{\rm H}$, such that $A(r_{\rm H})=B(r_{\rm H})=0$, and an essential singularity in $r=0$;
    \item they are both positive, smooth, and monotonic increasing functions for $r>r_{\rm H}$.
\end{enumerate}

The tetrad fields \eqref{eq:proper_tetrad} for the metric under study reads
\begin{align}
{\bf e}_{\hat t} &= \frac{{\bf e}_t}{\sqrt{A(r)}}\,,\ {\bf e}_{\hat r}=\sqrt{B(r)}\,{\bf e}_r\,,\ {\bf e}_{\hat \theta}= \frac{{\bf e}_\theta}{r}\,,\ {\bf e}_{\hat \varphi} = \frac{{\bf e}_\varphi}{r\sin\theta}\,.
\end{align}
The nonvanishing components of the Einstein tensor are
\begin{subequations}
\begin{align}
G_{\hat t\hat t}&=\frac{1-B-r B'}{r^2}\,, \\
G_{\hat r\hat r}&=\frac{A (B-1)+r A'B}{A r^2}\,, \\
G_{\hat\theta\hat\theta}&=\frac{2 A \left(rB A''+A B'\right)+A A' \left(r B'+2 B\right)-rB \left(A'\right)^2}{4r A^2 }\,, \\
G_{\hat\varphi\hat\varphi}&=G_{\hat\theta\hat\theta}\,,
\end{align}
\end{subequations}
where the prime denotes the derivative with respect to $r$. Additionally, the Ricci scalar is given by 
\begin{align}\label{eq:R}
R=&\ \frac{r^2B  \left(A'\right)^2-rA A' \left(r B'+4 B\right)}{2 r^2A^2 } \notag\\
&-\frac{r^2B  A''+2 rA  B'+2 A (B-1)}{r^2A }
\,.
\end{align}
Therefore, Eqs. \eqref{eq:LOC_SUM1} and \eqref{eq:LOC_SUM2} read 
\begin{subequations}
\begin{align}
&\frac{(2 A-r A')(A W B'+2 A B W'-B W A')}{A}+\frac{4A(B-1)W }{r} \notag \\
&=2 rB  W A''\,, \label{eq:BH-FE1}  \\
&2 B (r W A''+2 A r W''+6 A W')+2 A B' \left(r W'+W\right) \notag \\
&+A' (r W B'+4 B r W'+6 B W)+\frac{4 A (B-1) W}{r}\notag\\
&=\frac{B r W (A')^2}{A}\,.
\label{eq:BH-FE2}
\end{align}    
\end{subequations}
Equations~\eqref{eq:BH-FE1} and \eqref{eq:BH-FE2} represent a system of coupled differential equations involving both the metric and the scalar field $W(r)$.
We note that in both equations, $A(r)$ appears up to the second derivative, whereas $B(r)$ only up to the first. Therefore, in principle, one could assume for convenience the functional form of $A(r)$ and then try to solve the system for $B(r)$ and $W(r)$. 
However, this approach reveals to be very challenging when searching for analytical solutions. In the next section, we will present some strategies to overcome the intricate geometrical structure of the nonlocal field equations.

After solving the system \eqref{eq:BH-FE1}-\eqref{eq:BH-FE2}, one can then infer the functional form of the auxiliary scalar fields from Eqs.~\eqref{eq:X}--\eqref{eq:V}. In light of Eq.~\eqref{eq:BH-metric}, the d'Alembert operator \eqref{eq:box-operator} becomes
\begin{equation}
   \Box u(r)= \frac{1}{2} \left(\frac{B A'}{A}+B'+\frac{4 B}{r}\right)u' +B u''\,. \\
\end{equation}
This can be used to obtain $X(r)$ and $Y(r)$ from Eqs.~\eqref{eq:X} and \eqref{eq:Y}\footnote{Note that Eq.~\eqref{eq:Y} reads as
$\Box Y=g^{rr}[X'(r)]^2= B(r)[X'(r)]^2$.}, respectively, after calculating the scalar curvature as in \eqref{eq:R}.
Moreover, Eq.~\eqref{eq:U} can be recast as
\begin{equation}
\nabla_\mu(\nabla^\mu U+2V\nabla^\mu X)=0\,\Longrightarrow\ U'=-2VX'\,.
    \label{eq:U_1}
\end{equation}
Instead, from Eq.~\eqref{eq:W} we have
\begin{equation}\label{eq:f(r)}
f(r)=W(r)-U(r)-1\,,
\end{equation} 
so that, we can write  
\begin{equation}
\frac{\dd f}{\dd Y}=\frac{f'}{Y'}=\frac{W'-U'}{Y'}.
\label{eq:f'(Y)}
\end{equation}
Using Eqs.~\eqref{eq:U_1} and \eqref{eq:f'(Y)}, Eq.~\eqref{eq:V} becomes
\begin{equation}
\Box V=\left(\frac{W'+2VX'}{Y'}\right)R\,, 
\label{eq:V_1}
\end{equation}
which first allows one to determine $V(r)$ and then $U(r)$ from Eq.~\eqref{eq:U_1}. Next, $f(r)$ is given by Eq.~\eqref{eq:f(r)}. Finally, inverting $Y(r)$, one finds $r(Y)$ that can be inserted back into $f(r)$ to obtain the distortion function $f(Y)$. It is worth noting that all integration constants should be determined through the asymptotic flatness condition.

In the following, we investigate three classes of solutions arising from different ansatz on the metric functions. For consistency, we verify that the obtained solutions satisfy Eqs.~\eqref{eq:FE_LOC_t}--\eqref{eq:FE_LOC_phi} and also the original field equations \eqref{eq:FE} (or, equivalently, Eqs.~\eqref{eq:original-FE_t}--\eqref{eq:original-FE_phi}).

\subsection{First case: $A(r)=B(r)$}
\label{sec:I-case}
The simplest scenario is to assume that $-g_{tt}=1/g_{rr}$. In this case, Eqs.~\eqref{eq:BH-FE1} and \eqref{eq:BH-FE2} reduce to 
\begin{subequations}
\begin{align}
   & 2 A W' \left(2 A-r A'\right)+\frac{4 (A-1) A W}{r}=2 A r W A'' \,, \label{eq:first_case_1}\\
   & r W A''+A' (3 W'r+4 W)+2 A r W''+\frac{2 (A-1) W}{r}\notag\\
   &+6 A W'=0\,. \label{eq:first_case_2}
\end{align}
\end{subequations}
Although this case appears to be quite simple, determining an analytical solution is not trivial. However, we can handle the above field equations by looking for a perturbative correction to the Schwarzschild metric:
\begin{equation}
    A(r)=1-\frac{2}{r}-\alpha\, h(r)\,,
    \label{eq:fisrt_case_ansatz}
\end{equation}
where $|\alpha|\ll 1$, and $h(r)$ is a generic function to be determined.
Thus, from Eq.~\eqref{eq:first_case_1}, we have 
\begin{equation}
    W(r)=w_1 \exp \left[\int\frac{2 \alpha  h-\alpha r^2  h''}{\alpha r^2h'-2 \alpha r h+2 r-6}\dd r\right],
    \label{eq:first_case_W}
\end{equation}
with $w_1$ being an integration constant. Plugging Eq.~\eqref{eq:first_case_W} into Eq.~\eqref{eq:first_case_2}, at the first-order in $\alpha$ we find 
\begin{align}
   &r\left[r (6 - 5 r + r^2)h'''+(30 - 26 r + 5 r^2)h''\right] \nonumber \\ 
   &+ 2 (12 - 7 r + r^2)h'-2 h (r-4)=0\,,
\end{align}
whose general solution is
\begin{equation}
    h(r)=\frac{h_1}{r}+\frac{3r-r^2-2}{r}\left[h_2+\frac{h_3}{2}\ln\left(\frac{2-r}{r}\right)\right]+h_3\,,
\end{equation}
where $h_1$, $h_2$, and $h_3$ are constants of integration. The asymptotic flatness condition implies necessarily $h_2=h_3=0$, so we finally obtain $h(r)=h_1/r$, which can be reabsorbed in the unperturbed term of Eq.~\eqref{eq:fisrt_case_ansatz}, yielding finally the Schwarzschild solution.

In this case, the auxiliary scalar fields are all vanishing (i.e., $X(r)=Y(r)=U(r)=V(r)=0$), except for $W(r)=1$. Thus, the distortion function is $f(r)=f(Y)=0$ throughout the entire spacetime. 

In the most general case, $f(Y)\neq0$ for $r\in(r_{\rm H},\infty)$, while $R(r)$ and all auxiliary scalar fields vanish for $r\to\infty$, as required by the asymptotic flatness condition. To recover GR at infinity, namely $f(Y)\to0$, we also must impose $W(r)\rightarrow 1$ for $r\rightarrow \infty$. 
In the following calculations, we set all integration constants accordingly. 

\subsection{Second case: $A(r)=1-\frac{2}{r}$}
\label{sec:II-case}
As a second scenario, we study the case where $A(r)$ is chosen equal to the $tt$ component of the Schwarzschild metric, while we treat $B(r)$ perturbatively (cf. Eq. \eqref{eq:fisrt_case_ansatz}):
\begin{equation}
    B(r)=1-\frac{2}{r}-\alpha\, h(r)\,.
\end{equation}
Therefore, the integral solution of Eq.~\eqref{eq:LOC_SUM1} is
\begin{equation}
   \ln W(r)=\int\frac{\alpha  \left[2 h \left(3r-r^2 -3\right)-r \left(r^2-5 r+6\right) h'\right]}{2 \left(r^2-5 r+6\right) (2-r+\alpha  rh)}\dd r,
\end{equation}
which can be substituted into Eq.~\eqref{eq:LOC_SUM2} to obtain
\begin{equation}
    h(r)=\frac{r-2}{2 r(r-3)^2}\left\{(3-2 r) \left[h_1-2 h_2 \ln \left(\frac{2-r}{r}\right)\right]+6 h_2\right\},
\end{equation}
where $h_1$ and $h_2$ are integration constants. We note that an essential singularity occurs at $r=3$. The only way to remove it is to set $h_1=h_2=0$, which leads to $h(r)=0$, thus recovering the Schwarzschild metric.

\subsection{Third case: $A(r)=1-\frac{2}{r}-\frac{\alpha}{r^n}$}
\label{sec:III-case}

\begin{table*}
\renewcommand{\arraystretch}{2.6}
\centering
\scalebox{0.86}{
\begin{tabular}{|c|c|c|c|}
\hline
\hline
 & $n=2$ & $n=3$ & $n=4$ \\ \hline
$A(r)$ & $1-\dfrac{2}{r}-\dfrac{\alpha }{r^2}$ & $1-\dfrac{2}{r}-\dfrac{\alpha }{r^3}$ & $1-\dfrac{2}{r}-\dfrac{\alpha }{r^4}$\\
$B(r)$ & $1-\dfrac{2}{r}-\alpha\left(\dfrac{2 r-1}{3 r^2}\right)$ & $1-\dfrac{2}{r}-\alpha\left(\dfrac{2 r^2+5 r-9}{9 r^3}\right)$ & $1-\dfrac{2}{r}-\dfrac{\alpha  \left(2 r^3+5 r^2+18 r-45\right)}{27 r^4}$\\
\hline
$X(r)$ & $\alpha\left[x_1\ln\left(\dfrac{r}{r-2}\right)-\dfrac{1}{r}\right]$ & $\alpha  \left[x_1 \ln \left(\dfrac{r}{r-2}\right)-\dfrac{r+3}{3 r^2}\right]$ & $\alpha  \left[x_1 \ln \left(\dfrac{r}{r-2}\right)-\dfrac{r^2+3 r+9}{9 r^3}\right]$\\
$Y(r)$ & $\alpha  \ln \left(\dfrac{r}{r-2}\right)$ & $\alpha  \ln \left(\dfrac{r}{r-2}\right)$ & $\alpha  \ln \left(\dfrac{r}{r-2}\right)$ \\ 
$U(r)$ & 0 & 0 & 0\\  
$V(r)$ & $\alpha  \left[\dfrac{r-1}{12 r^2}+v_1 \ln \left(\dfrac{r}{r-2}\right)\right]$ & $\alpha  \left[\dfrac{r^3+5 r^2+4 r-18}{108 r^4}+v_1 \ln \left(\dfrac{r}{r-2}\right)\right]$ & $\alpha  \left[\dfrac{5 r^5+25 r^4+110 r^3+180 r^2+81 r-1215}{4860 r^6}+v_1 \ln \left(\dfrac{r}{r-2}\right)\right]$ \\ \hline  
$f(Y)$ & $\dfrac{\alpha}{6}   \left(e^{-Y/\alpha}-1\right)$ & $\dfrac{\alpha }{36}  \left(8 e^{-Y/\alpha }-3 e^{-2 Y/\alpha}-5\right)$ & $\dfrac{\alpha}{216} \left(e^{Y/\alpha }-1\right) \left(24 e^{-2 Y/\alpha}-9 e^{-3 Y/\alpha }-19 e^{-Y/\alpha }\right)$\\ 
\hline
\hline
\end{tabular}}
\caption{Summary of the nonlocal gravity BH solutions  discussed in Sec.~\ref{sec:III-case}, for $ n=2,3,4$.}
\label{tab:summary_fields}
\end{table*}

Let us now search for BH solutions beyond the Schwarzschild metric. For this purpose, we consider an inverse power-law correction for the $g_{tt}$ function and a perturbative correction for $g_{rr}$ as (cf. Eq. \eqref{eq:fisrt_case_ansatz}):
\begin{equation}
    B(r)=1-\frac{2}{r}-\alpha\, h(r)\,.
\end{equation}

In this case, Eq.~\eqref{eq:LOC_SUM1} provides
\begin{widetext}
\begin{equation}
    W(r)= w_1 \exp\left\{\int\frac{\alpha  \left[n^2 r^2-4 n^2 r+4 n^2+n r-2 n+2 r-2-\left(r^2-5 r+6\right) h' r^{n+1}-2 h \left(r^2-3 r+3\right) r^n\right]}{(r-2) \left[2 \alpha  h (r-3) r^{n+1}+2 \left(r^2-5 r+6\right) r^n+\alpha  r (n (r-2)+2)\right]}\dd r\right\}.
    \label{eq:third_case_W}
\end{equation}
\end{widetext}
Then, substituting the latter into Eq.~\eqref{eq:LOC_SUM2} and solving at first-order in $\alpha$, we find 
\begin{align}
    h(r)&=\frac{[n (r-3) (r-2)+4 r-9]}{r^n (r-3)^2}+\frac{h_1 [7 r-2(r^2+3)]}{2 r (r-3)^2},
    \label{eq:h(r)}
\end{align}
where $h_1$ is an integration constant. 
We note that a second-order singularity arises at $r=3$, which would also be reflected in the scalar curvature. 
To eliminate it, we require that the numerator of Eq.~\eqref{eq:h(r)} and its first derivative with respect to $r$ are both vanishing at the singularity point. This allows us to fix the value of the integration constant as $h_1=-2\times 3^{1-n}$. 
Under this choice, it can be easily verified that Eq.~\eqref{eq:h(r)} admits the finite limit 
\begin{equation}
    \lim_{r\rightarrow 3}h(r)=\frac{n^2+n+4}{2\times 3^{n+1}}\,.
\end{equation}

Notably, for $n=1$, we obtain $h(r)=1/r$, which reduces to the Schwarzschild solution. Let us then consider the general case $n>1$, where the metric components are:
\begin{subequations}
\begin{align}
    A(r)&=1-\frac{2}{r}-\frac{\alpha }{r^n}\,, \label{eq:BH_solA}\\
    B(r)&=1-\frac{2}{r}+\frac{\alpha}{3^{n} r^{n+1}(r-3)^2} \left\{3^n r \Big{[}n (r-3) (r-2)\right.\notag\\
    &\left.+4 r-9\Big{]}-3 (r-2) (2 r-3) r^n\right\}.
    \label{eq:BH_solB}
\end{align}    
\end{subequations}
In this scenario, the condition $A(r)=B(r)=0$ provides, at the first-order perturbation in $\alpha$, the event horizon
\begin{equation}
    r_{\rm H}=2+\frac{\alpha}{2^{n-1}}\,.
    \label{eq:rH}
\end{equation}
Moreover, from Eq.~\eqref{eq:third_case_W} at first order in $\alpha$, we find 
\begin{equation}
    W(r)=1+\frac{\alpha  \left(3^{1-n}-r^{1-n}\right)}{3-r}\,.
    \label{eq:W_sol}
\end{equation}

Then, we calculate the Ricci scalar\footnote{Notice that Eq.~\eqref{eq:R_sol} does not possess singularities beyond the event horizon, as $\lim_{r\rightarrow 3} R(r)=3^{-n-3}n\alpha\left(5+6n-n^2\right)$.} from Eq.~\eqref{eq:R}:
\begin{align}
    &R(r)=\frac{3\alpha r^{-n-\frac{5}{2}}}{(r-3)^3} \Biggl{[}3 \left(7 n^2-13 n+4\right) r^{\frac{3}{2}}+n(n-1) r^{\frac{7}{2}}\notag\\
    &+4 n (3-2 n) r^{\frac{5}{2}} -\frac{4r^{n+\frac{3}{2}}}{3^{n-1}}+\frac{2r^{n+\frac{1}{2}}}{3^{n-2}} -18 (n-1)^2 r^{\frac{1}{2}}\Biggl{]}.
    \label{eq:R_sol}
\end{align}
The latter can be used to solve Eq.~\eqref{eq:X} in the perturbed form $X(r)=X_0(r)+\alpha X_1(r)$, where $X_0(r)$ and $X_1(r)$ satisfy Eq.~\eqref{eq:X} at zeroth and first orders in $\alpha$, respectively.
Thus, the complete solution is
\begin{equation}
X(r)=\alpha  \left[\frac{3^{2-n}-3 r^{1-n}}{3-r}+x_1 \ln \left(\frac{r}{r-2}\right)\right],
\label{eq:X_sol}
\end{equation}
where $x_1$ is an arbitrary constant.
The same approach is pursued to solve Eq.~\eqref{eq:Y}, which gives
\begin{equation} 
Y(r)=\alpha  \ln \left(\frac{r}{r-2}\right) .
\label{eq:Ysol1}
\end{equation}
Using Eqs.~\eqref{eq:W_sol}, \eqref{eq:R_sol}, and \eqref{eq:X_sol}, it can be shown that the solution to Eq.~\eqref{eq:V_1} takes the general form
\begin{equation}
    V(r)=\alpha  \left[\frac{\mathcal{P}_n(r)}{r^{2n-2}}+v_1 \ln \left(\frac{r}{r-2}\right)\right],
    \label{eq:V_sol}
\end{equation}
where $\mathcal{P}_n(r)$ is a polynomial\footnote{The explicit expression of $\mathcal{P}_n(r)$ depends on the value of $n$ and may be not easy to determine for a generic $n$.} of order $2n-3$, and $v_1$ is an arbitrary constant.
Then, Eq.~\eqref{eq:V_sol} can be combined with Eq.~\eqref{eq:U_1} to obtain
\begin{equation}
    U(r)=0\,.
    \label{eq:U_sol}
\end{equation}
We note that Eqs.~\eqref{eq:Ysol1} and \eqref{eq:U_sol} do not depend on $n$ at first order in $\alpha$. This is since Eq.~\eqref{eq:X} involves $X(r)^2$, which in this case contains only the first-order $\alpha$ correction (cf. Eq.~\eqref{eq:X_sol}). The same argument applies also to Eq.~\eqref{eq:U_1} due to Eqs.~\eqref{eq:X_sol} and \eqref{eq:V_sol}.
Also, it is worth remarking that the auxiliary fields given by Eqs.~\eqref{eq:X_sol}, \eqref{eq:Ysol1}, and \eqref{eq:V_sol} are all finite when evaluated at the event horizon \eqref{eq:rH} (see Appendix~\ref{app:rH}).

Moreover, from Eq.~\eqref{eq:f(r)}, we get 
\begin{equation}
f(r)=\frac{\alpha  \left(3^{1-n}-r^{1-n}\right)}{3-r}\,.
\end{equation}
Inverting Eq. \eqref{eq:Ysol1}, we obtain
\begin{equation}
    r=\frac{2}{1-e^{Y/\alpha}}\,,
    \label{eq:r-Y}
\end{equation}
which finally permits to obtain the distortion function
\begin{equation}\label{eq:fY}
f(Y)=\frac{\alpha  \left(e^{Y/\alpha }-1\right)}{e^{Y/\alpha }-3}  \left[3^{1-n}-\left(\frac{2}{e^{Y/\alpha }-1}+2\right)^{1-n}\right].
\end{equation}
Consistently, the Schwarzschild limit $f(Y)\rightarrow 0$ is recovered for $\alpha\rightarrow 0$, $Y>0$. The upper limit of the $Y$ domain is determined by imposing the condition $r(Y)\ge r_{\rm H}$ via Eq.~\eqref{eq:r-Y}, which gives:
\begin{equation} \label{eq:Y_max}
    Y \le \alpha  \ln \left(\frac{2^n+\alpha}{\alpha }\right),
\end{equation}
which implies $\alpha>0$.

Let us consider the case $n\in \mathbb{N}$ for simplicity. 
In Table~\ref{tab:summary_fields}, we summarize the complete sets of BH solutions for the first nontrivial cases, which occur for $n=2,3,4$.

\begin{figure*}
    \includegraphics[width=3.3in]{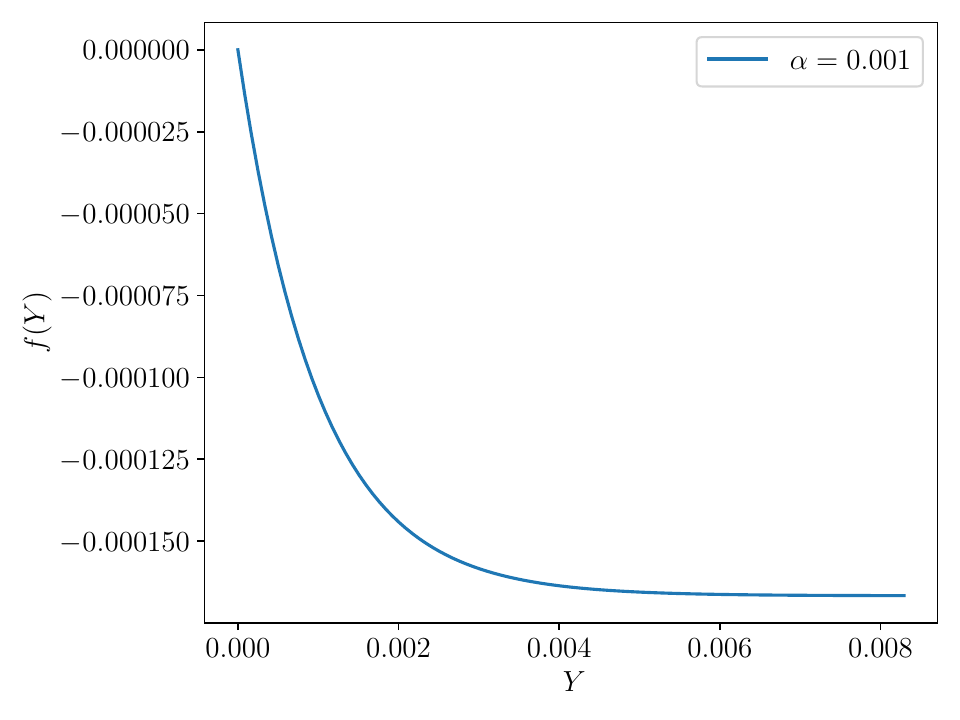}\quad 
    \includegraphics[width=3.3in]{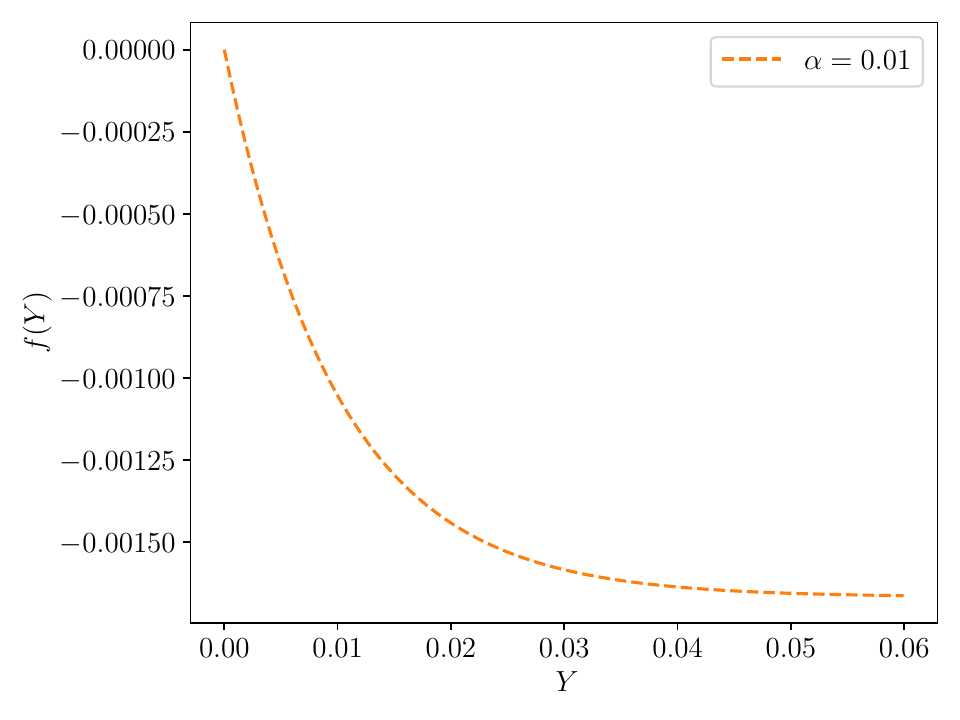} 
    \includegraphics[width=3.3in]{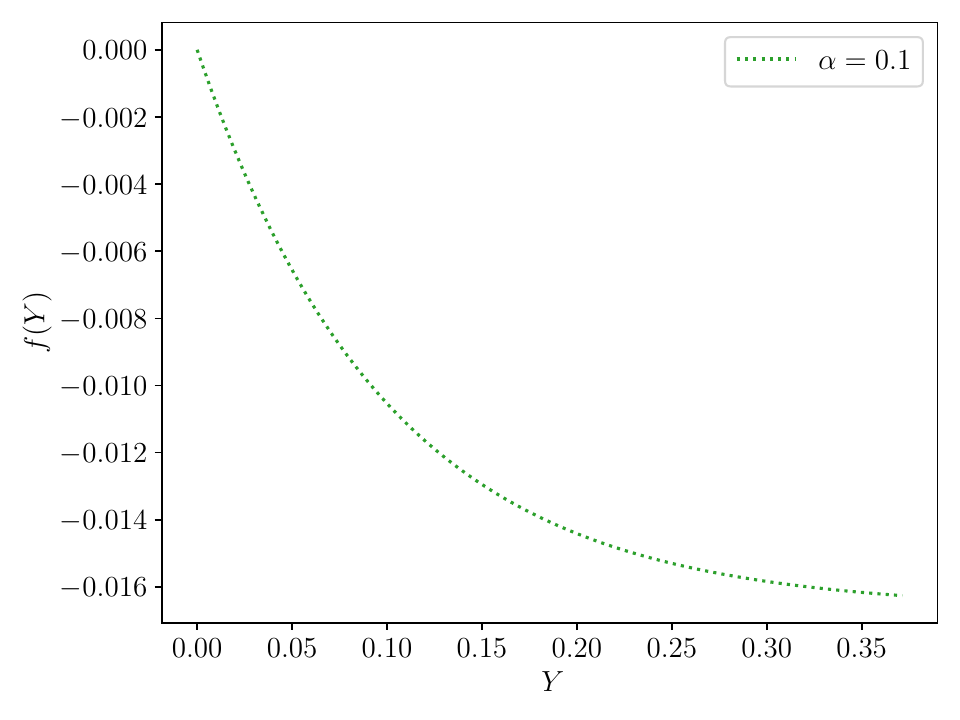}\quad 
    \includegraphics[width=3.3in]{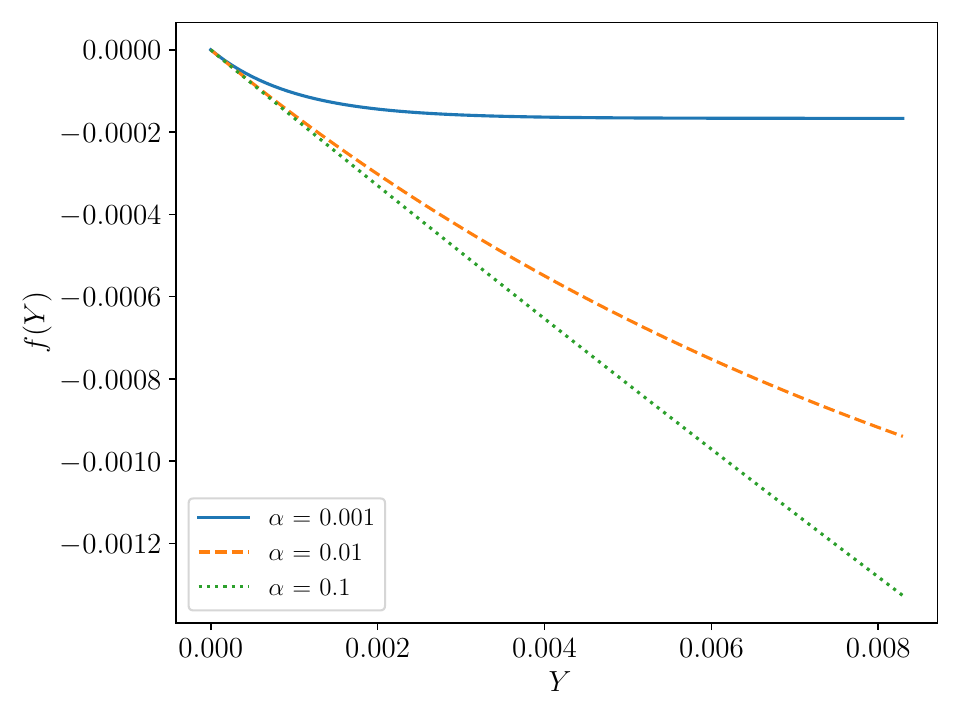}
    \caption{The first three panels display the distortion function $f(Y)$ for different values of the perturbation parameter $\alpha$ with fixed $n=2$, in the range $Y\in[0,Y(r_{\rm H})]$. The fourth panel compares the three $f(Y)$ profiles, where, to better highlight their differences, the horizontal axis is truncated at the value $Y(r_{\rm H})$ corresponding to $\alpha=0.001$.}
    \label{fig:distortion_function}
\end{figure*}

Furthermore, in Fig.~\ref{fig:distortion_function}, we show the profile of the distortion function in the case of $n=2$, for different values of $\alpha$. Specifically, we focus on $\alpha=(0.001,0.01,0.1)$, for which $Y(r_{\rm H})\simeq (0.008, 0.060, 0.369)$, respectively (see Eq.~\eqref{eq:Y_max}).  We observe that all curves converge to the Schwarzschild solution as $Y\rightarrow 0$, i.e., as $r\rightarrow \infty$. As expected, the deviations from the Schwarzschild prediction decrease as the $\alpha$ parameter becomes smaller.

\section{Black hole properties}
\label{sec:BH-properties}
To further characterize the configuration of the BH solution determined in Sec.~\ref{sec:III-case}, we calculate the following geometrical quantities \cite{DeFalco:2021klh}\footnote{To adapt the quantities listed in \cite{DeFalco:2021klh}, expressed using $g_{tt}=-e^{2\Phi(r)}$, we apply the transformation $\Phi(r)=\frac{1}{2}\ln[A(r)]$.
}:
\begin{itemize}
    \item the photon sphere radius, $r_{\rm ps}>r_{\rm H}$, obtained from
    \begin{equation}\label{eq:rps}
     r A'(r)-2 A(r)=0\,;   
    \end{equation}
    \item the critical impact parameter, $b_{\rm c}>r_{\rm H}$, which defines the radius of the compact object shadow:
    \begin{equation}\label{eq:bc}
     b_{\rm c}=\frac{r_{\rm ps}}{\sqrt{A(r_{\rm ps})}}\,;  
    \end{equation}
    \item the innermost stable circular orbit (ISCO) radius, $r_{\rm ISCO}>r_{\rm H}$, obtained by solving this equation in $r$:
    \begin{equation}\label{eq:rISCO}
     r \left(L^2+r^2\right) A'(r)-2 L^2 A(r)=0\,,   
    \end{equation}
    where $L$ is the conserved angular momentum along the test particle trajectory. Specifically, $r_{\rm ISCO}$ corresponds to the lowest value of $L$.
\end{itemize}

In the case of the BH solutions \eqref{eq:BH_solA}-\eqref{eq:BH_solB}, from Eqs.~\eqref{eq:rps} and \eqref{eq:bc}, respectively, we find
\begin{align}
r_{\rm ps}&=3 \left[1+\alpha \left(\frac{n+2}{2\times 3^n} \right)\right],   \\
b_{\rm c}&=3 \sqrt{3} \left(1+\alpha\frac{3^{1-n}}{2}   \right),
\end{align}
while Eq.~\eqref{eq:rISCO} is satisfied for
\begin{equation}
    r_{\rm ISCO}=6 \left(1-\alpha   \frac{3^{1-n}}{2^{n+2}}\right), \ \ L=2 \sqrt{3} \left[1+\alpha   \frac{(2 n+1)}{6^{n}}\right].
\end{equation}
In Table~\ref{tab:summary_radii}, we report the values of the above geometrical radii in the cases of $n=2,3,4$ to give an example. 

To quantify the corrections with respect to the Schwarzschild BH, let us consider the case $n=2$ and let us assume $\alpha=0.1$. This gives
\begin{equation}
r_{\rm H}=2.05\,, \ r_{ps}=3.07\,,\ \ b_c= 5.28\,, \ \
r_{\rm ISCO}=5.99\,.    
\end{equation}

From an experimental perspective, these represent small deviations from the Schwarzschild spacetime that would require both highly accurate observational data and the exploration of specific astrophysical phenomena capable of revealing them (see the discussion in \cite{DeFalco2023EPJC}).
\begin{table}
\renewcommand{\arraystretch}{2.5}
\centering
\begin{tabular}{|c|c|c|c|}
\hline
\hline
 & $n=2$ & $n=3$ & $n=4$ \\ \hline
$r_{\rm H}$ & $2+\dfrac{\alpha}{2}$ & $2+\dfrac{\alpha}{4}$ & $2+\dfrac{\alpha}{8}$\\ \hline
$r_{\rm ps}$ & $3+\dfrac{2\alpha}{3}$ & $3+\dfrac{5\alpha}{18}$ & $3+\dfrac{\alpha}{9}$\\ \hline
$b_{\rm c}$ & $\sqrt{3}\left(3+\dfrac{\alpha}{2}\right)$ & $\sqrt{3}\left(3+\dfrac{\alpha}{6}\right)$ & $\sqrt{3}\left(3+\dfrac{\alpha}{18}\right)$\\ \hline
$r_{\rm ISCO}$ & $6-\dfrac{\alpha}{8}$ & $6-\dfrac{\alpha}{48}$ & $6-\dfrac{\alpha}{288}$\\ \hline
\hline
\end{tabular}
\caption{Summary of the geometrical properties corresponding to the BH solutions reported in Table~\ref{tab:summary_fields}.}
\label{tab:summary_radii}
\end{table}

\section{Conclusion and perspectives}
\label{sec:end}
In this work, we revisited the Deser-Woodard model of nonlocal gravity using a new approach that maps the original field equations into a suitable tetrad frame. This method permits to significantly simplify the complex differential problem associated with the theory without altering the underlying gravitational framework. 

We thus considered the vacuum field equations of the nonlocal theory for a static and spherically symmetric spacetime. In particular, by analyzing different functional forms of the $g_{tt}$ and $g_{rr}$ metric functions, we demonstrated the existence of a new class of BH metrics. Specifically, these solutions arise from an inverse power-law correction to the Schwarzschild $g_{tt}$ metric function and a first-order perturbation to the Schwarzschild $g_{rr}$ component. We showed that the resulting BHs are free from essential singularities beyond the event horizon. At the same time, our analysis allowed us to determine the nonlocal auxiliary scalar fields of the theory through a cascade process, leading to the reconstruction of the distortion function and the ensuing gravitational action from which the new BH solutions originate. Finally, we obtained the analytical expressions of the photon sphere, the critical impact parameter (or shadow radius), and the ISCO radius within the first-order perturbation regime around the Schwarzschild BH spacetime.

The resulting class of BHs is parametrized by a real constant $n>1$. After deriving the analytical solutions for a generic $n$, we focused on the case $n\in\mathbb{N}$ and provided the explicit expressions for the first three nontrivial cases beyond the Schwarzschild solution, i.e., $n=2,3,4$. Additionally, we showed the magnitude of the corrections for typical small values of the perturbation parameter.

To the best of our knowledge, our findings introduce the first BH metrics beyond the Schwarzschild spacetime within the framework of the revised Deser-Woodard nonlocal gravity theory. Notably, the approach adopted in this study enabled us to work out the full theory analytically via a perturbative approach, despite the intricate geometric structure inherent to nonlocal gravity models. In general, our technique overcomes the commonly used methods in the literature, which typically impose \emph{a priori} requirements on the gravitational action or constraints on the forms of the auxiliary scalar fields.  On the other hand, going beyond the perturbative regime, analytical methods appear to yield limited results, whereas numerical techniques may represent the only viable approach.

The present work may serve as a foundation for discovering new solutions and further advancing the understanding of nonlocal theories and their connection to astrophysical compact objects. A promising future direction would be to benchmark the theoretical solutions against observational data, not only to validate the proposed models but also to impose stringent constraints on the theory itself, ultimately contributing to the development of a viable model of quantum gravity.

\section*{Acknowledgements}
R.D. acknowledges support from INFN -- Sezione di Roma 1, {\it esperimento} Euclid. V.D.F. thanks Gruppo Nazionale di Fisica Matematica of Istituto Nazionale di Alta Matematica for the support. V.D.F. acknowledges the support of INFN -- Sezione di Napoli, {\it iniziativa specifica} TEONGRAV. 

\appendix

\section{Original field equations}
\label{sec:full}
In this appendix, we highlight the advantages of our method using the recast field equations \eqref{eq:LOC_SUM} compared to the original ones \eqref{eq:FE}. In the latter case, the nonvanishing diagonal components are given by
\begin{subequations}
\begin{align}
&\left(G_{tt}+g_{tt}\Box+\nabla_t\nabla_t\right)W-\frac{1}{2}g_{tt}g^{rr}\mathcal{K}_{rr}=0\,,  \label{eq:original-FE_t}\\
&\left(G_{rr}+g_{rr}\Box+\nabla_r\nabla_r\right)W+\frac{1}{2}\mathcal{K}_{rr}=0 \,,\label{eq:original-FE_r} \\
&\left(G_{\varphi\varphi}+g_{\varphi\varphi}\Box+\nabla_\varphi\nabla_\varphi\right)W-\frac{1}{2}g_{\varphi\varphi}g^{rr}\mathcal{K}_{rr}=0\,. \label{eq:original-FE_phi}
\end{align}
\end{subequations}
Combining the above equations, we can reduce them to two independent ones, which are
\begin{subequations}
\begin{align}
    &(G_{tt}+G_{\varphi\varphi})W+(\nabla_t\nabla_t+\nabla_\varphi\nabla_\varphi)W+(g_{tt}+g_{\varphi\varphi})\Box W \notag \\ &=\frac{1}{2}(g_{tt}+g_{\varphi\varphi})g^{rr}\mathcal{K}_{rr}\,, \label{eq:original_SUM1}\\
    &(G_{tt}+G_{\varphi\varphi})W+(\nabla_r\nabla_r+\nabla_\varphi\nabla_\varphi)W+(g_{rr}+g_{\varphi\varphi})\Box W \notag \label{eq:original_SUM2}\\
    &=\frac{1}{2}(g_{\varphi\varphi}g^{rr}-1)\mathcal{K}_{rr}\,.
\end{align}
\end{subequations}
We immediately note that, in contrast with Eqs.~\eqref{eq:LOC_SUM1} and \eqref{eq:LOC_SUM2}, $\mathcal{K}_{rr}$ cannot be eliminated from Eqs.~\eqref{eq:original_SUM1} and \eqref{eq:original_SUM2}. Moreover, the latter equations generally appear more convoluted. This becomes evident when the metric functions \eqref{eq:BH-metric} are employed:
\begin{widetext}
\begin{subequations}
\begin{align}
    &\frac{A'(r W'+2 W)}{A}+r (A B+1) [X'(U'+V X')+V' Y']+4(1-A B) W'=\frac{2W (B-1)  (A B-1)}{B r}+2 A B r W'' \notag  \\
    &+ A B' (r W'+2 W) \label{eq:original_FE1}\,, \\
    &\frac{B r^3 W (A')^2}{A}=2 \left\{r \left[B r^2 W A''+A (1-B r^2) U' X'-A B r^2 V' Y'+A V (1-B r^2) \left(X'\right)^2+2 A B r^2 W''+A V' Y'\right]\right. \notag \\
    &\left.+A r^2 B' \left(r W'+W\right)+2 A \left(B r^2+2\right) W'\right\}+A' \left[r^3 W B'+2 \left(B r^3+r\right) W'+2 W \left(B r^2+2\right)\right]+\frac{4 A (B-1) W}{B r}\,. \label{eq:original_FE2}
\end{align}
\end{subequations}
\end{widetext}
Therefore, a direct comparison of Eqs.~\eqref{eq:original_FE1} and \eqref{eq:original_FE2} with Eqs.~\eqref{eq:BH-FE1} and \eqref{eq:BH-FE2}, respectively, underscores the great effectiveness of our approach.

\section{Behavior of the auxiliary fields at the event horizon}
\label{app:rH}

We report below the values assumed by the auxiliary scalar fields when evaluated at the event horizon \eqref{eq:rH} for a generic $n$. Specifically, from Eqs.~\eqref{eq:X_sol}, \eqref{eq:Ysol1} and \eqref{eq:V_sol}, we find, respectively,
\begin{align}
\lim_{r\to r_{\rm H}}X(r)&=\alpha  \left[3^{2-n}-\frac{3}{2^{n-1}} -x_1 \ln \left(\frac{\alpha}{2^{n}}  \right)\right],\\
\lim_{r\to r_{\rm H}}Y(r)&=\alpha  \ln \left(\frac{2^n}{\alpha }\right),\\
\lim_{r\to r_{\rm H}}V(r)&=\alpha\left[ \frac{\mathcal{P}_n(2)}{4^{n-1}}+v_1 \ln \left(\frac{2^n}{\alpha }\right)\right],
\end{align}
whereas $U(r)$ identically vanishes, as per Eq.~\eqref{eq:U_sol}.

\clearpage

\bibliography{references}

\begin{thebibliography}{53}%
\makeatletter
\providecommand \@ifxundefined [1]{%
 \@ifx{#1\undefined}
}%
\providecommand \@ifnum [1]{%
 \ifnum #1\expandafter \@firstoftwo
 \else \expandafter \@secondoftwo
 \fi
}%
\providecommand \@ifx [1]{%
 \ifx #1\expandafter \@firstoftwo
 \else \expandafter \@secondoftwo
 \fi
}%
\providecommand \natexlab [1]{#1}%
\providecommand \enquote  [1]{``#1''}%
\providecommand \bibnamefont  [1]{#1}%
\providecommand \bibfnamefont [1]{#1}%
\providecommand \citenamefont [1]{#1}%
\providecommand \href@noop [0]{\@secondoftwo}%
\providecommand \href [0]{\begingroup \@sanitize@url \@href}%
\providecommand \@href[1]{\@@startlink{#1}\@@href}%
\providecommand \@@href[1]{\endgroup#1\@@endlink}%
\providecommand \@sanitize@url [0]{\catcode `\\12\catcode `\$12\catcode
  `\&12\catcode `\#12\catcode `\^12\catcode `\_12\catcode `\%12\relax}%
\providecommand \@@startlink[1]{}%
\providecommand \@@endlink[0]{}%
\providecommand \url  [0]{\begingroup\@sanitize@url \@url }%
\providecommand \@url [1]{\endgroup\@href {#1}{\urlprefix }}%
\providecommand \urlprefix  [0]{URL }%
\providecommand \Eprint [0]{\href }%
\providecommand \doibase [0]{http://dx.doi.org/}%
\providecommand \selectlanguage [0]{\@gobble}%
\providecommand \bibinfo  [0]{\@secondoftwo}%
\providecommand \bibfield  [0]{\@secondoftwo}%
\providecommand \translation [1]{[#1]}%
\providecommand \BibitemOpen [0]{}%
\providecommand \bibitemStop [0]{}%
\providecommand \bibitemNoStop [0]{.\EOS\space}%
\providecommand \EOS [0]{\spacefactor3000\relax}%
\providecommand \BibitemShut  [1]{\csname bibitem#1\endcsname}%
\let\auto@bib@innerbib\@empty
\bibitem [{\citenamefont {Padmanabhan}(2002)}]{Padmanabhan:2001ev}%
  \BibitemOpen
  \bibfield  {author} {\bibinfo {author} {\bibfnamefont {T.}~\bibnamefont
  {Padmanabhan}},\ }\href {\doibase 10.1088/0264-9381/19/13/312} {\bibfield
  {journal} {\bibinfo  {journal} {Class. Quant. Grav.}\ }\textbf {\bibinfo
  {volume} {19}},\ \bibinfo {pages} {3551} (\bibinfo {year} {2002})},\ \Eprint
  {http://arxiv.org/abs/gr-qc/0110046} {arXiv:gr-qc/0110046} \BibitemShut
  {NoStop}%
\bibitem [{\citenamefont {Ashtekar}\ and\ \citenamefont
  {Lewandowski}(2004)}]{Ashtekar:2004eh}%
  \BibitemOpen
  \bibfield  {author} {\bibinfo {author} {\bibfnamefont {A.}~\bibnamefont
  {Ashtekar}}\ and\ \bibinfo {author} {\bibfnamefont {J.}~\bibnamefont
  {Lewandowski}},\ }\href {\doibase 10.1088/0264-9381/21/15/R01} {\bibfield
  {journal} {\bibinfo  {journal} {Class. Quant. Grav.}\ }\textbf {\bibinfo
  {volume} {21}},\ \bibinfo {pages} {R53} (\bibinfo {year} {2004})},\ \Eprint
  {http://arxiv.org/abs/gr-qc/0404018} {arXiv:gr-qc/0404018} \BibitemShut
  {NoStop}%
\bibitem [{\citenamefont {Will}(2014)}]{Will:2014kxa}%
  \BibitemOpen
  \bibfield  {author} {\bibinfo {author} {\bibfnamefont {C.~M.}\ \bibnamefont
  {Will}},\ }\href {\doibase 10.12942/lrr-2014-4} {\bibfield  {journal}
  {\bibinfo  {journal} {Living Rev. Rel.}\ }\textbf {\bibinfo {volume} {17}},\
  \bibinfo {pages} {4} (\bibinfo {year} {2014})},\ \Eprint
  {http://arxiv.org/abs/1403.7377} {arXiv:1403.7377 [gr-qc]} \BibitemShut
  {NoStop}%
\bibitem [{\citenamefont {Abbott}\ \emph {et~al.}(2016)\citenamefont {Abbott}
  \emph {et~al.}}]{LIGOScientific:2016aoc}%
  \BibitemOpen
  \bibfield  {author} {\bibinfo {author} {\bibfnamefont {B.~P.}\ \bibnamefont
  {Abbott}} \emph {et~al.} (\bibinfo {collaboration} {LIGO Scientific,
  Virgo}),\ }\href {\doibase 10.1103/PhysRevLett.116.061102} {\bibfield
  {journal} {\bibinfo  {journal} {Phys. Rev. Lett.}\ }\textbf {\bibinfo
  {volume} {116}},\ \bibinfo {pages} {061102} (\bibinfo {year} {2016})},\
  \Eprint {http://arxiv.org/abs/1602.03837} {arXiv:1602.03837 [gr-qc]}
  \BibitemShut {NoStop}%
\bibitem [{\citenamefont {Akiyama}\ \emph {et~al.}(2019)\citenamefont {Akiyama}
  \emph {et~al.}}]{EventHorizonTelescope:2019dse}%
  \BibitemOpen
  \bibfield  {author} {\bibinfo {author} {\bibfnamefont {K.}~\bibnamefont
  {Akiyama}} \emph {et~al.} (\bibinfo {collaboration} {Event Horizon
  Telescope}),\ }\href {\doibase 10.3847/2041-8213/ab0ec7} {\bibfield
  {journal} {\bibinfo  {journal} {Astrophys. J. Lett.}\ }\textbf {\bibinfo
  {volume} {875}},\ \bibinfo {pages} {L1} (\bibinfo {year} {2019})},\ \Eprint
  {http://arxiv.org/abs/1906.11238} {arXiv:1906.11238 [astro-ph.GA]}
  \BibitemShut {NoStop}%
\bibitem [{\citenamefont {Ishak}(2019)}]{Ishak:2018his}%
  \BibitemOpen
  \bibfield  {author} {\bibinfo {author} {\bibfnamefont {M.}~\bibnamefont
  {Ishak}},\ }\href {\doibase 10.1007/s41114-018-0017-4} {\bibfield  {journal}
  {\bibinfo  {journal} {Living Rev. Rel.}\ }\textbf {\bibinfo {volume} {22}},\
  \bibinfo {pages} {1} (\bibinfo {year} {2019})},\ \Eprint
  {http://arxiv.org/abs/1806.10122} {arXiv:1806.10122 [astro-ph.CO]}
  \BibitemShut {NoStop}%
\bibitem [{\citenamefont {Barack}\ \emph {et~al.}(2019)\citenamefont {Barack}
  \emph {et~al.}}]{Barack:2018yly}%
  \BibitemOpen
  \bibfield  {author} {\bibinfo {author} {\bibfnamefont {L.}~\bibnamefont
  {Barack}} \emph {et~al.},\ }\href {\doibase 10.1088/1361-6382/ab0587}
  {\bibfield  {journal} {\bibinfo  {journal} {Class. Quant. Grav.}\ }\textbf
  {\bibinfo {volume} {36}},\ \bibinfo {pages} {143001} (\bibinfo {year}
  {2019})},\ \Eprint {http://arxiv.org/abs/1806.05195} {arXiv:1806.05195
  [gr-qc]} \BibitemShut {NoStop}%
\bibitem [{\citenamefont {D'Agostino}\ and\ \citenamefont
  {Nunes}(2023)}]{DAgostino:2023cgx}%
  \BibitemOpen
  \bibfield  {author} {\bibinfo {author} {\bibfnamefont {R.}~\bibnamefont
  {D'Agostino}}\ and\ \bibinfo {author} {\bibfnamefont {R.~C.}\ \bibnamefont
  {Nunes}},\ }\href {\doibase 10.1103/PhysRevD.108.023523} {\bibfield
  {journal} {\bibinfo  {journal} {Phys. Rev. D}\ }\textbf {\bibinfo {volume}
  {108}},\ \bibinfo {pages} {023523} (\bibinfo {year} {2023})},\ \Eprint
  {http://arxiv.org/abs/2307.13464} {arXiv:2307.13464 [astro-ph.CO]}
  \BibitemShut {NoStop}%
\bibitem [{\citenamefont {Riess}\ \emph {et~al.}(1998)\citenamefont {Riess}
  \emph {et~al.}}]{SupernovaSearchTeam:1998fmf}%
  \BibitemOpen
  \bibfield  {author} {\bibinfo {author} {\bibfnamefont {A.~G.}\ \bibnamefont
  {Riess}} \emph {et~al.} (\bibinfo {collaboration} {Supernova Search Team}),\
  }\href {\doibase 10.1086/300499} {\bibfield  {journal} {\bibinfo  {journal}
  {Astron. J.}\ }\textbf {\bibinfo {volume} {116}},\ \bibinfo {pages} {1009}
  (\bibinfo {year} {1998})},\ \Eprint {http://arxiv.org/abs/astro-ph/9805201}
  {arXiv:astro-ph/9805201} \BibitemShut {NoStop}%
\bibitem [{\citenamefont {Perlmutter}\ \emph {et~al.}(1999)\citenamefont
  {Perlmutter} \emph {et~al.}}]{SupernovaCosmologyProject:1998vns}%
  \BibitemOpen
  \bibfield  {author} {\bibinfo {author} {\bibfnamefont {S.}~\bibnamefont
  {Perlmutter}} \emph {et~al.} (\bibinfo {collaboration} {Supernova Cosmology
  Project}),\ }\href {\doibase 10.1086/307221} {\bibfield  {journal} {\bibinfo
  {journal} {Astrophys. J.}\ }\textbf {\bibinfo {volume} {517}},\ \bibinfo
  {pages} {565} (\bibinfo {year} {1999})},\ \Eprint
  {http://arxiv.org/abs/astro-ph/9812133} {arXiv:astro-ph/9812133} \BibitemShut
  {NoStop}%
\bibitem [{\citenamefont {Peebles}\ and\ \citenamefont
  {Ratra}(2003)}]{Peebles:2002gy}%
  \BibitemOpen
  \bibfield  {author} {\bibinfo {author} {\bibfnamefont {P.~J.~E.}\
  \bibnamefont {Peebles}}\ and\ \bibinfo {author} {\bibfnamefont
  {B.}~\bibnamefont {Ratra}},\ }\href {\doibase 10.1103/RevModPhys.75.559}
  {\bibfield  {journal} {\bibinfo  {journal} {Rev. Mod. Phys.}\ }\textbf
  {\bibinfo {volume} {75}},\ \bibinfo {pages} {559} (\bibinfo {year} {2003})},\
  \Eprint {http://arxiv.org/abs/astro-ph/0207347} {arXiv:astro-ph/0207347}
  \BibitemShut {NoStop}%
\bibitem [{\citenamefont {Aghanim}\ \emph {et~al.}(2020)\citenamefont {Aghanim}
  \emph {et~al.}}]{Planck:2018vyg}%
  \BibitemOpen
  \bibfield  {author} {\bibinfo {author} {\bibfnamefont {N.}~\bibnamefont
  {Aghanim}} \emph {et~al.} (\bibinfo {collaboration} {Planck}),\ }\href
  {\doibase 10.1051/0004-6361/201833910} {\bibfield  {journal} {\bibinfo
  {journal} {Astron. Astrophys.}\ }\textbf {\bibinfo {volume} {641}},\ \bibinfo
  {pages} {A6} (\bibinfo {year} {2020})},\ \bibinfo {note} {[Erratum:
  Astron.Astrophys. 652, C4 (2021)]},\ \Eprint
  {http://arxiv.org/abs/1807.06209} {arXiv:1807.06209 [astro-ph.CO]}
  \BibitemShut {NoStop}%
\bibitem [{\citenamefont {D'Agostino}(2019)}]{DAgostino:2019wko}%
  \BibitemOpen
  \bibfield  {author} {\bibinfo {author} {\bibfnamefont {R.}~\bibnamefont
  {D'Agostino}},\ }\href {\doibase 10.1103/PhysRevD.99.103524} {\bibfield
  {journal} {\bibinfo  {journal} {Phys. Rev. D}\ }\textbf {\bibinfo {volume}
  {99}},\ \bibinfo {pages} {103524} (\bibinfo {year} {2019})},\ \Eprint
  {http://arxiv.org/abs/1903.03836} {arXiv:1903.03836 [gr-qc]} \BibitemShut
  {NoStop}%
\bibitem [{\citenamefont {Weinberg}(1989)}]{Weinberg:1988cp}%
  \BibitemOpen
  \bibfield  {author} {\bibinfo {author} {\bibfnamefont {S.}~\bibnamefont
  {Weinberg}},\ }\href {\doibase 10.1103/RevModPhys.61.1} {\bibfield  {journal}
  {\bibinfo  {journal} {Rev. Mod. Phys.}\ }\textbf {\bibinfo {volume} {61}},\
  \bibinfo {pages} {1} (\bibinfo {year} {1989})}\BibitemShut {NoStop}%
\bibitem [{\citenamefont {Carroll}(2001)}]{Carroll:2000fy}%
  \BibitemOpen
  \bibfield  {author} {\bibinfo {author} {\bibfnamefont {S.~M.}\ \bibnamefont
  {Carroll}},\ }\href {\doibase 10.12942/lrr-2001-1} {\bibfield  {journal}
  {\bibinfo  {journal} {Living Rev. Rel.}\ }\textbf {\bibinfo {volume} {4}},\
  \bibinfo {pages} {1} (\bibinfo {year} {2001})},\ \Eprint
  {http://arxiv.org/abs/astro-ph/0004075} {arXiv:astro-ph/0004075} \BibitemShut
  {NoStop}%
\bibitem [{\citenamefont {D'Agostino}\ \emph {et~al.}(2022)\citenamefont
  {D'Agostino}, \citenamefont {Luongo},\ and\ \citenamefont
  {Muccino}}]{DAgostino:2022fcx}%
  \BibitemOpen
  \bibfield  {author} {\bibinfo {author} {\bibfnamefont {R.}~\bibnamefont
  {D'Agostino}}, \bibinfo {author} {\bibfnamefont {O.}~\bibnamefont {Luongo}},
  \ and\ \bibinfo {author} {\bibfnamefont {M.}~\bibnamefont {Muccino}},\ }\href
  {\doibase 10.1088/1361-6382/ac8af2} {\bibfield  {journal} {\bibinfo
  {journal} {Class. Quant. Grav.}\ }\textbf {\bibinfo {volume} {39}},\ \bibinfo
  {pages} {195014} (\bibinfo {year} {2022})},\ \Eprint
  {http://arxiv.org/abs/2204.02190} {arXiv:2204.02190 [gr-qc]} \BibitemShut
  {NoStop}%
\bibitem [{\citenamefont {Silvestri}\ and\ \citenamefont
  {Trodden}(2009)}]{Silvestri:2009hh}%
  \BibitemOpen
  \bibfield  {author} {\bibinfo {author} {\bibfnamefont {A.}~\bibnamefont
  {Silvestri}}\ and\ \bibinfo {author} {\bibfnamefont {M.}~\bibnamefont
  {Trodden}},\ }\href {\doibase 10.1088/0034-4885/72/9/096901} {\bibfield
  {journal} {\bibinfo  {journal} {Rept. Prog. Phys.}\ }\textbf {\bibinfo
  {volume} {72}},\ \bibinfo {pages} {096901} (\bibinfo {year} {2009})},\
  \Eprint {http://arxiv.org/abs/0904.0024} {arXiv:0904.0024 [astro-ph.CO]}
  \BibitemShut {NoStop}%
\bibitem [{\citenamefont {De~Felice}\ and\ \citenamefont
  {Tsujikawa}(2010)}]{DeFelice:2010aj}%
  \BibitemOpen
  \bibfield  {author} {\bibinfo {author} {\bibfnamefont {A.}~\bibnamefont
  {De~Felice}}\ and\ \bibinfo {author} {\bibfnamefont {S.}~\bibnamefont
  {Tsujikawa}},\ }\href {\doibase 10.12942/lrr-2010-3} {\bibfield  {journal}
  {\bibinfo  {journal} {Living Rev. Rel.}\ }\textbf {\bibinfo {volume} {13}},\
  \bibinfo {pages} {3} (\bibinfo {year} {2010})},\ \Eprint
  {http://arxiv.org/abs/1002.4928} {arXiv:1002.4928 [gr-qc]} \BibitemShut
  {NoStop}%
\bibitem [{\citenamefont {Clifton}\ \emph {et~al.}(2012)\citenamefont
  {Clifton}, \citenamefont {Ferreira}, \citenamefont {Padilla},\ and\
  \citenamefont {Skordis}}]{Clifton:2011jh}%
  \BibitemOpen
  \bibfield  {author} {\bibinfo {author} {\bibfnamefont {T.}~\bibnamefont
  {Clifton}}, \bibinfo {author} {\bibfnamefont {P.~G.}\ \bibnamefont
  {Ferreira}}, \bibinfo {author} {\bibfnamefont {A.}~\bibnamefont {Padilla}}, \
  and\ \bibinfo {author} {\bibfnamefont {C.}~\bibnamefont {Skordis}},\ }\href
  {\doibase 10.1016/j.physrep.2012.01.001} {\bibfield  {journal} {\bibinfo
  {journal} {Phys. Rept.}\ }\textbf {\bibinfo {volume} {513}},\ \bibinfo
  {pages} {1} (\bibinfo {year} {2012})},\ \Eprint
  {http://arxiv.org/abs/1106.2476} {arXiv:1106.2476 [astro-ph.CO]} \BibitemShut
  {NoStop}%
\bibitem [{\citenamefont {Joyce}\ \emph {et~al.}(2015)\citenamefont {Joyce},
  \citenamefont {Jain}, \citenamefont {Khoury},\ and\ \citenamefont
  {Trodden}}]{Joyce:2014kja}%
  \BibitemOpen
  \bibfield  {author} {\bibinfo {author} {\bibfnamefont {A.}~\bibnamefont
  {Joyce}}, \bibinfo {author} {\bibfnamefont {B.}~\bibnamefont {Jain}},
  \bibinfo {author} {\bibfnamefont {J.}~\bibnamefont {Khoury}}, \ and\ \bibinfo
  {author} {\bibfnamefont {M.}~\bibnamefont {Trodden}},\ }\href {\doibase
  10.1016/j.physrep.2014.12.002} {\bibfield  {journal} {\bibinfo  {journal}
  {Phys. Rept.}\ }\textbf {\bibinfo {volume} {568}},\ \bibinfo {pages} {1}
  (\bibinfo {year} {2015})},\ \Eprint {http://arxiv.org/abs/1407.0059}
  {arXiv:1407.0059 [astro-ph.CO]} \BibitemShut {NoStop}%
\bibitem [{\citenamefont {Nojiri}\ \emph {et~al.}(2017)\citenamefont {Nojiri},
  \citenamefont {Odintsov},\ and\ \citenamefont {Oikonomou}}]{Nojiri:2017ncd}%
  \BibitemOpen
  \bibfield  {author} {\bibinfo {author} {\bibfnamefont {S.}~\bibnamefont
  {Nojiri}}, \bibinfo {author} {\bibfnamefont {S.~D.}\ \bibnamefont
  {Odintsov}}, \ and\ \bibinfo {author} {\bibfnamefont {V.~K.}\ \bibnamefont
  {Oikonomou}},\ }\href {\doibase 10.1016/j.physrep.2017.06.001} {\bibfield
  {journal} {\bibinfo  {journal} {Phys. Rept.}\ }\textbf {\bibinfo {volume}
  {692}},\ \bibinfo {pages} {1} (\bibinfo {year} {2017})},\ \Eprint
  {http://arxiv.org/abs/1705.11098} {arXiv:1705.11098 [gr-qc]} \BibitemShut
  {NoStop}%
\bibitem [{\citenamefont {Capozziello}\ \emph {et~al.}(2019)\citenamefont
  {Capozziello}, \citenamefont {D'Agostino},\ and\ \citenamefont
  {Luongo}}]{Capozziello:2019cav}%
  \BibitemOpen
  \bibfield  {author} {\bibinfo {author} {\bibfnamefont {S.}~\bibnamefont
  {Capozziello}}, \bibinfo {author} {\bibfnamefont {R.}~\bibnamefont
  {D'Agostino}}, \ and\ \bibinfo {author} {\bibfnamefont {O.}~\bibnamefont
  {Luongo}},\ }\href {\doibase 10.1142/S0218271819300167} {\bibfield  {journal}
  {\bibinfo  {journal} {Int. J. Mod. Phys. D}\ }\textbf {\bibinfo {volume}
  {28}},\ \bibinfo {pages} {1930016} (\bibinfo {year} {2019})},\ \Eprint
  {http://arxiv.org/abs/1904.01427} {arXiv:1904.01427 [gr-qc]} \BibitemShut
  {NoStop}%
\bibitem [{\citenamefont {D'Agostino}\ and\ \citenamefont
  {Nunes}(2019)}]{DAgostino:2019hvh}%
  \BibitemOpen
  \bibfield  {author} {\bibinfo {author} {\bibfnamefont {R.}~\bibnamefont
  {D'Agostino}}\ and\ \bibinfo {author} {\bibfnamefont {R.~C.}\ \bibnamefont
  {Nunes}},\ }\href {\doibase 10.1103/PhysRevD.100.044041} {\bibfield
  {journal} {\bibinfo  {journal} {Phys. Rev. D}\ }\textbf {\bibinfo {volume}
  {100}},\ \bibinfo {pages} {044041} (\bibinfo {year} {2019})},\ \Eprint
  {http://arxiv.org/abs/1907.05516} {arXiv:1907.05516 [gr-qc]} \BibitemShut
  {NoStop}%
\bibitem [{\citenamefont {D'Agostino}\ and\ \citenamefont
  {Nunes}(2022)}]{DAgostino:2022tdk}%
  \BibitemOpen
  \bibfield  {author} {\bibinfo {author} {\bibfnamefont {R.}~\bibnamefont
  {D'Agostino}}\ and\ \bibinfo {author} {\bibfnamefont {R.~C.}\ \bibnamefont
  {Nunes}},\ }\href {\doibase 10.1103/PhysRevD.106.124053} {\bibfield
  {journal} {\bibinfo  {journal} {Phys. Rev. D}\ }\textbf {\bibinfo {volume}
  {106}},\ \bibinfo {pages} {124053} (\bibinfo {year} {2022})},\ \Eprint
  {http://arxiv.org/abs/2210.11935} {arXiv:2210.11935 [gr-qc]} \BibitemShut
  {NoStop}%
\bibitem [{\citenamefont {D'Agostino}\ \emph {et~al.}(2024)\citenamefont
  {D'Agostino}, \citenamefont {Luongo},\ and\ \citenamefont
  {Mancini}}]{DAgostino:2024ymo}%
  \BibitemOpen
  \bibfield  {author} {\bibinfo {author} {\bibfnamefont {R.}~\bibnamefont
  {D'Agostino}}, \bibinfo {author} {\bibfnamefont {O.}~\bibnamefont {Luongo}},
  \ and\ \bibinfo {author} {\bibfnamefont {S.}~\bibnamefont {Mancini}},\ }\href
  {\doibase 10.1140/epjc/s10052-024-13440-y} {\bibfield  {journal} {\bibinfo
  {journal} {Eur. Phys. J. C}\ }\textbf {\bibinfo {volume} {84}},\ \bibinfo
  {pages} {1060} (\bibinfo {year} {2024})},\ \Eprint
  {http://arxiv.org/abs/2403.06819} {arXiv:2403.06819 [gr-qc]} \BibitemShut
  {NoStop}%
\bibitem [{\citenamefont {Arkani-Hamed}\ \emph {et~al.}(2002)\citenamefont
  {Arkani-Hamed}, \citenamefont {Dimopoulos}, \citenamefont {Dvali},\ and\
  \citenamefont {Gabadadze}}]{Arkani-Hamed:2002ukf}%
  \BibitemOpen
  \bibfield  {author} {\bibinfo {author} {\bibfnamefont {N.}~\bibnamefont
  {Arkani-Hamed}}, \bibinfo {author} {\bibfnamefont {S.}~\bibnamefont
  {Dimopoulos}}, \bibinfo {author} {\bibfnamefont {G.}~\bibnamefont {Dvali}}, \
  and\ \bibinfo {author} {\bibfnamefont {G.}~\bibnamefont {Gabadadze}},\
  }\href@noop {} {\  (\bibinfo {year} {2002})},\ \Eprint
  {http://arxiv.org/abs/hep-th/0209227} {arXiv:hep-th/0209227} \BibitemShut
  {NoStop}%
\bibitem [{\citenamefont {Nojiri}\ and\ \citenamefont
  {Odintsov}(2008)}]{Nojiri:2007uq}%
  \BibitemOpen
  \bibfield  {author} {\bibinfo {author} {\bibfnamefont {S.}~\bibnamefont
  {Nojiri}}\ and\ \bibinfo {author} {\bibfnamefont {S.~D.}\ \bibnamefont
  {Odintsov}},\ }\href {\doibase 10.1016/j.physletb.2007.12.001} {\bibfield
  {journal} {\bibinfo  {journal} {Phys. Lett. B}\ }\textbf {\bibinfo {volume}
  {659}},\ \bibinfo {pages} {821} (\bibinfo {year} {2008})},\ \Eprint
  {http://arxiv.org/abs/0708.0924} {arXiv:0708.0924 [hep-th]} \BibitemShut
  {NoStop}%
\bibitem [{\citenamefont {Calcagni}\ \emph {et~al.}(2007)\citenamefont
  {Calcagni}, \citenamefont {Montobbio},\ and\ \citenamefont
  {Nardelli}}]{Calcagni:2007ru}%
  \BibitemOpen
  \bibfield  {author} {\bibinfo {author} {\bibfnamefont {G.}~\bibnamefont
  {Calcagni}}, \bibinfo {author} {\bibfnamefont {M.}~\bibnamefont {Montobbio}},
  \ and\ \bibinfo {author} {\bibfnamefont {G.}~\bibnamefont {Nardelli}},\
  }\href {\doibase 10.1103/PhysRevD.76.126001} {\bibfield  {journal} {\bibinfo
  {journal} {Phys. Rev. D}\ }\textbf {\bibinfo {volume} {76}},\ \bibinfo
  {pages} {126001} (\bibinfo {year} {2007})},\ \Eprint
  {http://arxiv.org/abs/0705.3043} {arXiv:0705.3043 [hep-th]} \BibitemShut
  {NoStop}%
\bibitem [{\citenamefont {Biswas}\ \emph {et~al.}(2012)\citenamefont {Biswas},
  \citenamefont {Gerwick}, \citenamefont {Koivisto},\ and\ \citenamefont
  {Mazumdar}}]{Biswas:2011ar}%
  \BibitemOpen
  \bibfield  {author} {\bibinfo {author} {\bibfnamefont {T.}~\bibnamefont
  {Biswas}}, \bibinfo {author} {\bibfnamefont {E.}~\bibnamefont {Gerwick}},
  \bibinfo {author} {\bibfnamefont {T.}~\bibnamefont {Koivisto}}, \ and\
  \bibinfo {author} {\bibfnamefont {A.}~\bibnamefont {Mazumdar}},\ }\href
  {\doibase 10.1103/PhysRevLett.108.031101} {\bibfield  {journal} {\bibinfo
  {journal} {Phys. Rev. Lett.}\ }\textbf {\bibinfo {volume} {108}},\ \bibinfo
  {pages} {031101} (\bibinfo {year} {2012})},\ \Eprint
  {http://arxiv.org/abs/1110.5249} {arXiv:1110.5249 [gr-qc]} \BibitemShut
  {NoStop}%
\bibitem [{\citenamefont {Park}\ and\ \citenamefont
  {Dodelson}(2013)}]{Park:2012cp}%
  \BibitemOpen
  \bibfield  {author} {\bibinfo {author} {\bibfnamefont {S.}~\bibnamefont
  {Park}}\ and\ \bibinfo {author} {\bibfnamefont {S.}~\bibnamefont
  {Dodelson}},\ }\href {\doibase 10.1103/PhysRevD.87.024003} {\bibfield
  {journal} {\bibinfo  {journal} {Phys. Rev. D}\ }\textbf {\bibinfo {volume}
  {87}},\ \bibinfo {pages} {024003} (\bibinfo {year} {2013})},\ \Eprint
  {http://arxiv.org/abs/1209.0836} {arXiv:1209.0836 [astro-ph.CO]} \BibitemShut
  {NoStop}%
\bibitem [{\citenamefont {Maggiore}\ and\ \citenamefont
  {Mancarella}(2014)}]{Maggiore:2014sia}%
  \BibitemOpen
  \bibfield  {author} {\bibinfo {author} {\bibfnamefont {M.}~\bibnamefont
  {Maggiore}}\ and\ \bibinfo {author} {\bibfnamefont {M.}~\bibnamefont
  {Mancarella}},\ }\href {\doibase 10.1103/PhysRevD.90.023005} {\bibfield
  {journal} {\bibinfo  {journal} {Phys. Rev. D}\ }\textbf {\bibinfo {volume}
  {90}},\ \bibinfo {pages} {023005} (\bibinfo {year} {2014})},\ \Eprint
  {http://arxiv.org/abs/1402.0448} {arXiv:1402.0448 [hep-th]} \BibitemShut
  {NoStop}%
\bibitem [{\citenamefont {Dirian}\ \emph {et~al.}(2014)\citenamefont {Dirian},
  \citenamefont {Foffa}, \citenamefont {Khosravi}, \citenamefont {Kunz},\ and\
  \citenamefont {Maggiore}}]{Dirian:2014ara}%
  \BibitemOpen
  \bibfield  {author} {\bibinfo {author} {\bibfnamefont {Y.}~\bibnamefont
  {Dirian}}, \bibinfo {author} {\bibfnamefont {S.}~\bibnamefont {Foffa}},
  \bibinfo {author} {\bibfnamefont {N.}~\bibnamefont {Khosravi}}, \bibinfo
  {author} {\bibfnamefont {M.}~\bibnamefont {Kunz}}, \ and\ \bibinfo {author}
  {\bibfnamefont {M.}~\bibnamefont {Maggiore}},\ }\href {\doibase
  10.1088/1475-7516/2014/06/033} {\bibfield  {journal} {\bibinfo  {journal}
  {JCAP}\ }\textbf {\bibinfo {volume} {06}},\ \bibinfo {pages} {033} (\bibinfo
  {year} {2014})},\ \Eprint {http://arxiv.org/abs/1403.6068} {arXiv:1403.6068
  [astro-ph.CO]} \BibitemShut {NoStop}%
\bibitem [{\citenamefont {Capozziello}\ \emph {et~al.}(2022)\citenamefont
  {Capozziello}, \citenamefont {D'Agostino},\ and\ \citenamefont
  {Luongo}}]{Capozziello:2022rac}%
  \BibitemOpen
  \bibfield  {author} {\bibinfo {author} {\bibfnamefont {S.}~\bibnamefont
  {Capozziello}}, \bibinfo {author} {\bibfnamefont {R.}~\bibnamefont
  {D'Agostino}}, \ and\ \bibinfo {author} {\bibfnamefont {O.}~\bibnamefont
  {Luongo}},\ }\href {\doibase 10.1016/j.physletb.2022.137475} {\bibfield
  {journal} {\bibinfo  {journal} {Phys. Lett. B}\ }\textbf {\bibinfo {volume}
  {834}},\ \bibinfo {pages} {137475} (\bibinfo {year} {2022})},\ \Eprint
  {http://arxiv.org/abs/2207.01276} {arXiv:2207.01276 [gr-qc]} \BibitemShut
  {NoStop}%
\bibitem [{\citenamefont {Deffayet}\ and\ \citenamefont
  {Woodard}(2024)}]{Deffayet:2024ciu}%
  \BibitemOpen
  \bibfield  {author} {\bibinfo {author} {\bibfnamefont {C.}~\bibnamefont
  {Deffayet}}\ and\ \bibinfo {author} {\bibfnamefont {R.~P.}\ \bibnamefont
  {Woodard}},\ }\href {\doibase 10.1088/1475-7516/2024/05/042} {\bibfield
  {journal} {\bibinfo  {journal} {JCAP}\ }\textbf {\bibinfo {volume} {05}},\
  \bibinfo {pages} {042} (\bibinfo {year} {2024})},\ \Eprint
  {http://arxiv.org/abs/2402.11716} {arXiv:2402.11716 [gr-qc]} \BibitemShut
  {NoStop}%
\bibitem [{\citenamefont {Modesto}(2012)}]{Modesto:2011kw}%
  \BibitemOpen
  \bibfield  {author} {\bibinfo {author} {\bibfnamefont {L.}~\bibnamefont
  {Modesto}},\ }\href {\doibase 10.1103/PhysRevD.86.044005} {\bibfield
  {journal} {\bibinfo  {journal} {Phys. Rev. D}\ }\textbf {\bibinfo {volume}
  {86}},\ \bibinfo {pages} {044005} (\bibinfo {year} {2012})},\ \Eprint
  {http://arxiv.org/abs/1107.2403} {arXiv:1107.2403 [hep-th]} \BibitemShut
  {NoStop}%
\bibitem [{\citenamefont {Modesto}\ and\ \citenamefont
  {Rachwa\l{}}(2017)}]{Modesto:2017sdr}%
  \BibitemOpen
  \bibfield  {author} {\bibinfo {author} {\bibfnamefont {L.}~\bibnamefont
  {Modesto}}\ and\ \bibinfo {author} {\bibfnamefont {L.}~\bibnamefont
  {Rachwa\l{}}},\ }\href {\doibase 10.1142/S0218271817300208} {\bibfield
  {journal} {\bibinfo  {journal} {Int. J. Mod. Phys. D}\ }\textbf {\bibinfo
  {volume} {26}},\ \bibinfo {pages} {1730020} (\bibinfo {year}
  {2017})}\BibitemShut {NoStop}%
\bibitem [{\citenamefont {Buoninfante}\ \emph {et~al.}(2019)\citenamefont
  {Buoninfante}, \citenamefont {Lambiase},\ and\ \citenamefont
  {Mazumdar}}]{Buoninfante:2018mre}%
  \BibitemOpen
  \bibfield  {author} {\bibinfo {author} {\bibfnamefont {L.}~\bibnamefont
  {Buoninfante}}, \bibinfo {author} {\bibfnamefont {G.}~\bibnamefont
  {Lambiase}}, \ and\ \bibinfo {author} {\bibfnamefont {A.}~\bibnamefont
  {Mazumdar}},\ }\href {\doibase 10.1016/j.nuclphysb.2019.114646} {\bibfield
  {journal} {\bibinfo  {journal} {Nucl. Phys. B}\ }\textbf {\bibinfo {volume}
  {944}},\ \bibinfo {pages} {114646} (\bibinfo {year} {2019})},\ \Eprint
  {http://arxiv.org/abs/1805.03559} {arXiv:1805.03559 [hep-th]} \BibitemShut
  {NoStop}%
\bibitem [{\citenamefont {Deser}\ and\ \citenamefont
  {Woodard}(2007)}]{Deser:2007jk}%
  \BibitemOpen
  \bibfield  {author} {\bibinfo {author} {\bibfnamefont {S.}~\bibnamefont
  {Deser}}\ and\ \bibinfo {author} {\bibfnamefont {R.~P.}\ \bibnamefont
  {Woodard}},\ }\href {\doibase 10.1103/PhysRevLett.99.111301} {\bibfield
  {journal} {\bibinfo  {journal} {Phys. Rev. Lett.}\ }\textbf {\bibinfo
  {volume} {99}},\ \bibinfo {pages} {111301} (\bibinfo {year} {2007})},\
  \Eprint {http://arxiv.org/abs/0706.2151} {arXiv:0706.2151 [astro-ph]}
  \BibitemShut {NoStop}%
\bibitem [{\citenamefont {Belgacem}\ \emph {et~al.}(2019)\citenamefont
  {Belgacem}, \citenamefont {Finke}, \citenamefont {Frassino},\ and\
  \citenamefont {Maggiore}}]{Belgacem:2018wtb}%
  \BibitemOpen
  \bibfield  {author} {\bibinfo {author} {\bibfnamefont {E.}~\bibnamefont
  {Belgacem}}, \bibinfo {author} {\bibfnamefont {A.}~\bibnamefont {Finke}},
  \bibinfo {author} {\bibfnamefont {A.}~\bibnamefont {Frassino}}, \ and\
  \bibinfo {author} {\bibfnamefont {M.}~\bibnamefont {Maggiore}},\ }\href
  {\doibase 10.1088/1475-7516/2019/02/035} {\bibfield  {journal} {\bibinfo
  {journal} {JCAP}\ }\textbf {\bibinfo {volume} {02}},\ \bibinfo {pages} {035}
  (\bibinfo {year} {2019})},\ \Eprint {http://arxiv.org/abs/1812.11181}
  {arXiv:1812.11181 [gr-qc]} \BibitemShut {NoStop}%
\bibitem [{\citenamefont {Deser}\ and\ \citenamefont
  {Woodard}(2019)}]{Deser:2019lmm}%
  \BibitemOpen
  \bibfield  {author} {\bibinfo {author} {\bibfnamefont {S.}~\bibnamefont
  {Deser}}\ and\ \bibinfo {author} {\bibfnamefont {R.~P.}\ \bibnamefont
  {Woodard}},\ }\href {\doibase 10.1088/1475-7516/2019/06/034} {\bibfield
  {journal} {\bibinfo  {journal} {JCAP}\ }\textbf {\bibinfo {volume} {06}},\
  \bibinfo {pages} {034} (\bibinfo {year} {2019})},\ \Eprint
  {http://arxiv.org/abs/1902.08075} {arXiv:1902.08075 [gr-qc]} \BibitemShut
  {NoStop}%
\bibitem [{\citenamefont {Ding}\ and\ \citenamefont
  {Deng}(2019)}]{Ding:2019rlp}%
  \BibitemOpen
  \bibfield  {author} {\bibinfo {author} {\bibfnamefont {J.-C.}\ \bibnamefont
  {Ding}}\ and\ \bibinfo {author} {\bibfnamefont {J.-B.}\ \bibnamefont
  {Deng}},\ }\href {\doibase 10.1088/1475-7516/2019/12/054} {\bibfield
  {journal} {\bibinfo  {journal} {JCAP}\ }\textbf {\bibinfo {volume} {12}},\
  \bibinfo {pages} {054} (\bibinfo {year} {2019})},\ \Eprint
  {http://arxiv.org/abs/1908.11223} {arXiv:1908.11223 [astro-ph.CO]}
  \BibitemShut {NoStop}%
\bibitem [{\citenamefont {Chen}\ \emph {et~al.}(2019)\citenamefont {Chen},
  \citenamefont {Chen},\ and\ \citenamefont {Park}}]{Chen:2019wlu}%
  \BibitemOpen
  \bibfield  {author} {\bibinfo {author} {\bibfnamefont {C.-Y.}\ \bibnamefont
  {Chen}}, \bibinfo {author} {\bibfnamefont {P.}~\bibnamefont {Chen}}, \ and\
  \bibinfo {author} {\bibfnamefont {S.}~\bibnamefont {Park}},\ }\href {\doibase
  10.1016/j.physletb.2019.07.024} {\bibfield  {journal} {\bibinfo  {journal}
  {Phys. Lett. B}\ }\textbf {\bibinfo {volume} {796}},\ \bibinfo {pages} {112}
  (\bibinfo {year} {2019})},\ \Eprint {http://arxiv.org/abs/1905.04557}
  {arXiv:1905.04557 [gr-qc]} \BibitemShut {NoStop}%
\bibitem [{\citenamefont {Jackson}\ and\ \citenamefont
  {Bufalo}(2022)}]{Jackson:2021mgw}%
  \BibitemOpen
  \bibfield  {author} {\bibinfo {author} {\bibfnamefont {D.}~\bibnamefont
  {Jackson}}\ and\ \bibinfo {author} {\bibfnamefont {R.}~\bibnamefont
  {Bufalo}},\ }\href {\doibase 10.1088/1475-7516/2022/05/043} {\bibfield
  {journal} {\bibinfo  {journal} {JCAP}\ }\textbf {\bibinfo {volume} {05}},\
  \bibinfo {pages} {043} (\bibinfo {year} {2022})},\ \Eprint
  {http://arxiv.org/abs/2110.10008} {arXiv:2110.10008 [gr-qc]} \BibitemShut
  {NoStop}%
\bibitem [{\citenamefont {Capozziello}\ and\ \citenamefont
  {D'Agostino}(2023)}]{Capozziello:2023ccw}%
  \BibitemOpen
  \bibfield  {author} {\bibinfo {author} {\bibfnamefont {S.}~\bibnamefont
  {Capozziello}}\ and\ \bibinfo {author} {\bibfnamefont {R.}~\bibnamefont
  {D'Agostino}},\ }\href {\doibase 10.1016/j.dark.2023.101346} {\bibfield
  {journal} {\bibinfo  {journal} {Phys. Dark Univ.}\ }\textbf {\bibinfo
  {volume} {42}},\ \bibinfo {pages} {101346} (\bibinfo {year} {2023})},\
  \Eprint {http://arxiv.org/abs/2310.03136} {arXiv:2310.03136 [gr-qc]}
  \BibitemShut {NoStop}%
\bibitem [{\citenamefont {Jackson}\ and\ \citenamefont
  {Bufalo}(2023)}]{Jackson:2023faq}%
  \BibitemOpen
  \bibfield  {author} {\bibinfo {author} {\bibfnamefont {D.}~\bibnamefont
  {Jackson}}\ and\ \bibinfo {author} {\bibfnamefont {R.}~\bibnamefont
  {Bufalo}},\ }\href {\doibase 10.1088/1475-7516/2023/05/010} {\bibfield
  {journal} {\bibinfo  {journal} {JCAP}\ }\textbf {\bibinfo {volume} {05}},\
  \bibinfo {pages} {010} (\bibinfo {year} {2023})},\ \Eprint
  {http://arxiv.org/abs/2303.17552} {arXiv:2303.17552 [gr-qc]} \BibitemShut
  {NoStop}%
\bibitem [{\citenamefont {Chen}\ and\ \citenamefont
  {Park}(2021)}]{Chen:2021pxd}%
  \BibitemOpen
  \bibfield  {author} {\bibinfo {author} {\bibfnamefont {C.-Y.}\ \bibnamefont
  {Chen}}\ and\ \bibinfo {author} {\bibfnamefont {S.}~\bibnamefont {Park}},\
  }\href {\doibase 10.1103/PhysRevD.103.064029} {\bibfield  {journal} {\bibinfo
   {journal} {Phys. Rev. D}\ }\textbf {\bibinfo {volume} {103}},\ \bibinfo
  {pages} {064029} (\bibinfo {year} {2021})},\ \Eprint
  {http://arxiv.org/abs/2101.06600} {arXiv:2101.06600 [gr-qc]} \BibitemShut
  {NoStop}%
\bibitem [{\citenamefont {Kehagias}\ and\ \citenamefont
  {Maggiore}(2014)}]{Kehagias:2014sda}%
  \BibitemOpen
  \bibfield  {author} {\bibinfo {author} {\bibfnamefont {A.}~\bibnamefont
  {Kehagias}}\ and\ \bibinfo {author} {\bibfnamefont {M.}~\bibnamefont
  {Maggiore}},\ }\href {\doibase 10.1007/JHEP08(2014)029} {\bibfield  {journal}
  {\bibinfo  {journal} {JHEP}\ }\textbf {\bibinfo {volume} {08}},\ \bibinfo
  {pages} {029} (\bibinfo {year} {2014})},\ \Eprint
  {http://arxiv.org/abs/1401.8289} {arXiv:1401.8289 [hep-th]} \BibitemShut
  {NoStop}%
\bibitem [{\citenamefont {Calcagni}\ \emph {et~al.}(2018)\citenamefont
  {Calcagni}, \citenamefont {Modesto},\ and\ \citenamefont
  {Myung}}]{Calcagni:2018pro}%
  \BibitemOpen
  \bibfield  {author} {\bibinfo {author} {\bibfnamefont {G.}~\bibnamefont
  {Calcagni}}, \bibinfo {author} {\bibfnamefont {L.}~\bibnamefont {Modesto}}, \
  and\ \bibinfo {author} {\bibfnamefont {Y.~S.}\ \bibnamefont {Myung}},\ }\href
  {\doibase 10.1016/j.physletb.2018.06.041} {\bibfield  {journal} {\bibinfo
  {journal} {Phys. Lett. B}\ }\textbf {\bibinfo {volume} {783}},\ \bibinfo
  {pages} {19} (\bibinfo {year} {2018})},\ \Eprint
  {http://arxiv.org/abs/1803.08388} {arXiv:1803.08388 [gr-qc]} \BibitemShut
  {NoStop}%
\bibitem [{\citenamefont {Kumar}\ \emph {et~al.}(2020)\citenamefont {Kumar},
  \citenamefont {Panda},\ and\ \citenamefont {Patel}}]{Kumar:2019uwi}%
  \BibitemOpen
  \bibfield  {author} {\bibinfo {author} {\bibfnamefont {U.}~\bibnamefont
  {Kumar}}, \bibinfo {author} {\bibfnamefont {S.}~\bibnamefont {Panda}}, \ and\
  \bibinfo {author} {\bibfnamefont {A.}~\bibnamefont {Patel}},\ }\href
  {\doibase 10.1140/epjc/s10052-020-8182-5} {\bibfield  {journal} {\bibinfo
  {journal} {Eur. Phys. J. C}\ }\textbf {\bibinfo {volume} {80}},\ \bibinfo
  {pages} {614} (\bibinfo {year} {2020})},\ \Eprint
  {http://arxiv.org/abs/1906.11714} {arXiv:1906.11714 [gr-qc]} \BibitemShut
  {NoStop}%
\bibitem [{\citenamefont {Buoninfante}\ \emph {et~al.}(2024)\citenamefont
  {Buoninfante}, \citenamefont {Giacchini},\ and\ \citenamefont
  {de~Paula~Netto}}]{Buoninfante:2022ild}%
  \BibitemOpen
  \bibfield  {author} {\bibinfo {author} {\bibfnamefont {L.}~\bibnamefont
  {Buoninfante}}, \bibinfo {author} {\bibfnamefont {B.~L.}\ \bibnamefont
  {Giacchini}}, \ and\ \bibinfo {author} {\bibfnamefont {T.}~\bibnamefont
  {de~Paula~Netto}},\ }\enquote {\bibinfo {title} {{Black Holes in Non-local
  Gravity}},}\ \ (\bibinfo {year} {2024})\ \Eprint
  {http://arxiv.org/abs/2211.03497} {arXiv:2211.03497 [gr-qc]} \BibitemShut
  {NoStop}%
\bibitem [{\citenamefont {Morris}\ and\ \citenamefont
  {Thorne}(1988)}]{Morris:1988cz}%
  \BibitemOpen
  \bibfield  {author} {\bibinfo {author} {\bibfnamefont {M.~S.}\ \bibnamefont
  {Morris}}\ and\ \bibinfo {author} {\bibfnamefont {K.~S.}\ \bibnamefont
  {Thorne}},\ }\href {\doibase 10.1119/1.15620} {\bibfield  {journal} {\bibinfo
   {journal} {Am. J. Phys.}\ }\textbf {\bibinfo {volume} {56}},\ \bibinfo
  {pages} {395} (\bibinfo {year} {1988})}\BibitemShut {NoStop}%
\bibitem [{\citenamefont {De~Falco}\ \emph {et~al.}(2021)\citenamefont
  {De~Falco}, \citenamefont {Battista}, \citenamefont {Capozziello},\ and\
  \citenamefont {De~Laurentis}}]{DeFalco:2021klh}%
  \BibitemOpen
  \bibfield  {author} {\bibinfo {author} {\bibfnamefont {V.}~\bibnamefont
  {De~Falco}}, \bibinfo {author} {\bibfnamefont {E.}~\bibnamefont {Battista}},
  \bibinfo {author} {\bibfnamefont {S.}~\bibnamefont {Capozziello}}, \ and\
  \bibinfo {author} {\bibfnamefont {M.}~\bibnamefont {De~Laurentis}},\ }\href
  {\doibase 10.1103/PhysRevD.103.044007} {\bibfield  {journal} {\bibinfo
  {journal} {Phys. Rev. D}\ }\textbf {\bibinfo {volume} {103}},\ \bibinfo
  {pages} {044007} (\bibinfo {year} {2021})},\ \Eprint
  {http://arxiv.org/abs/2101.04960} {arXiv:2101.04960 [gr-qc]} \BibitemShut
  {NoStop}%
\bibitem [{\citenamefont {De~Falco}\ \emph {et~al.}(2023)\citenamefont
  {De~Falco}, \citenamefont {Bajardi}, \citenamefont {D'Agostino},
  \citenamefont {Benetti},\ and\ \citenamefont
  {Capozziello}}]{DeFalco2023EPJC}%
  \BibitemOpen
  \bibfield  {author} {\bibinfo {author} {\bibfnamefont {V.}~\bibnamefont
  {De~Falco}}, \bibinfo {author} {\bibfnamefont {F.}~\bibnamefont {Bajardi}},
  \bibinfo {author} {\bibfnamefont {R.}~\bibnamefont {D'Agostino}}, \bibinfo
  {author} {\bibfnamefont {M.}~\bibnamefont {Benetti}}, \ and\ \bibinfo
  {author} {\bibfnamefont {S.}~\bibnamefont {Capozziello}},\ }\href {\doibase
  10.1140/epjc/s10052-023-11601-z} {\bibfield  {journal} {\bibinfo  {journal}
  {Eur. Phys. J. C}\ }\textbf {\bibinfo {volume} {83}},\ \bibinfo {pages} {456}
  (\bibinfo {year} {2023})},\ \Eprint {http://arxiv.org/abs/2305.04695}
  {arXiv:2305.04695 [gr-qc]} \BibitemShut {NoStop}%
\end{thebibliography}%

\end{document}